\documentclass[iop,twocolumn]{emulateapj}
\usepackage[english]{babel}
\usepackage[utf8x]{inputenc}
\usepackage{amsmath}
\usepackage{graphicx}
\usepackage{natbib}
\pdfoutput=1
\usepackage{color}
\usepackage[usenames,dvipsnames]{xcolor}
\usepackage{mathtools}
\usepackage{comment}
\usepackage{hyperref}

\newcommand{\hii}{H\,{\sc ii}}
\newcommand{\ci}{C\,{\sc i}}
\newcommand{\cii}{C\,{\sc ii}}
\newcommand{\oi}{O\,{\sc i}}
\newcommand{\oiii}{O\,{\sc iii}}
\newcommand{\nii}{N\,{\sc ii}}

\newcommand{\sii}{S\,{\sc ii}}
\newcommand{\mum}{\ensuremath{\,\mu\mbox{m}}}

\begin{document}

\title{The Survey of Lines in M\,31 (SLIM): Investigating the Origins of [\cii] Emission}

\author{M.~J.~Kapala\altaffilmark{1}, K.~Sandstrom\altaffilmark{1,}\altaffilmark{2}, B.~Groves\altaffilmark{1}, K.~Croxall\altaffilmark{3}, K.~Kreckel\altaffilmark{1}, J.~Dalcanton\altaffilmark{4}, A.~Leroy\altaffilmark{5} E.~Schinnerer\altaffilmark{1}, F.~Walter\altaffilmark{1}, M.~Fouesneau\altaffilmark{1}}

\altaffiltext{1}{Max Planck Institut f\"ur Astronomie, K\"onigstuhl 17, D-69117 Heidelberg, Germany; kapala@mpia.de}
\altaffiltext{2}{Steward Observatory, Arizona, USA}
\altaffiltext{3}{Department of Astronomy, The Ohio State University, 140 West 18th Avenue, Columbus, OH 43210, USA}
\altaffiltext{4}{Department of Astronomy, University of Washington, Box 351580, Seattle, WA 98195, USA}
\altaffiltext{5}{National Radio Astronomy Observatory, 520 Edgemont Road, Charlottesville, VA 22903, USA}

\begin{abstract}
The [\cii] 158\,\mum\ line is one of the strongest emission lines observed in star-forming galaxies, and has been empirically measured to correlate with the star formation rate (SFR) globally and on $\sim$\,kpc scales. However, due to the multi-phase origins of [\cii], one might expect this relation to break down at small scales. We investigate the origins of [\cii] emission by examining high spatial resolution observations of [\cii] in M\,31, with the Survey of Lines in M\,31 (SLIM). We present five $\sim700\times$700\,pc (3\arcmin$\times$3$\arcmin$) Fields mapping the [\cii] emission, H$\alpha$ emission, combined with ancillary infrared (IR) data. We spatially separate star-forming regions from diffuse gas and dust emission on $\sim$\,50\,pc scales. We find that the [\cii] -- SFR correlation holds even at these scales, although the relation typically has a flatter slope than found at larger ($\sim$\,kpc) scales. 
While the  H$\alpha$ emission in M\,31 is concentrated in the SFR regions, we find that a significant amount ($\sim$\,20--90\%) of the [\cii] emission comes from outside star-forming regions, and that the total IR (TIR) emission has the highest diffuse fraction of all SFR tracers. We find a weak correlation of the [\cii]/TIR to dust color in each Field, and find a large scale trend of increasing [\cii]/TIR with  galactocentric radius. The differences in the relative diffuse fractions of [\cii], H$\alpha$ and IR tracers are likely caused by a combination of energetic photon leakage from \hii\ regions and heating by the diffuse radiation field arising from older (B-star) stellar populations. However, we find that by averaging our measurements over $\sim$\,kpc scales, these effects are minimized, and the relation between [\cii] and SFR found in other nearby galaxy 
studies is retrieved.


\end{abstract}

\section{Introduction}

[\cii] 158\,\mum\ is one of the strongest emission lines observed in star-forming galaxies and is the dominant coolant of the neutral interstellar medium \citep{Luhman1998}. Its strength is due to the relatively high abundance of singly-ionized carbon in the interstellar medium (ISM) and the wide range of physical conditions (densities and temperatures) in which the fine-structure transition can be collisionally excited \citep[for a review, see][]{Goldsmith2012}. 
Due to its strength and long wavelength, the [\cii] line is relatively easy to observe even at high redshifts with sub-mm telescopes, such as ALMA \citep{Carilli2013}.  It can potentially be used to estimate the star formation rate (SFR) and, with additional information from other emission lines, to probe some of the  physical conditions of the ISM. 

Many recent studies attempted to empirically calibrate the relation between [\cii] and SFR, both on global and kpc scales. They all find approximately linear relationships between [\cii] and SFR with small scatter \citep[see e.g.,][]{deLooze2011, deLooze2014, Herrera2014}. The physical basis for this relation relies upon [\cii] being the dominant coolant of the neutral ISM heated by the UV photons from young stars, via the photoelectric process on dust grains \citep{Tielens1985, Wolfire1995}. 

One issue with the interpretation of the [\cii]--SFR relation, is that the [\cii] 158\,\mum\ emission line arises from multiple ISM phases. The ionization potential of carbon (11.26\,eV) is less than that of hydrogen (13.6\,eV), such that C$^+$ is also present in the neutral ISM, where the [\cii] line is typically excited via collisions with atomic hydrogen. 
At the other extreme, the ionization potential of C$^+$ is sufficiently high (24.4\,eV) that the ion is also found in the ionized ISM, where collisions with electrons dominate the line excitation. In addition, the C$^+$ species can also be found in molecular gas before the atom combines to \ci\ and CO \citep{Wolfire2010}. In each phase, the [\cii] line has a different sensitivity to the gas density (i.e.~different critical densities), and different relation with the heating radiation, and hence SFR. 

Not only do the energetics of [\cii] vary with ISM phase, but the timescale over which it measures SFR varies as well. In the ionized ISM, heating is by the photoionization of hydrogen caused by the extreme UV photons ($ >13.6$\,eV) from O-stars, and thus [\cii] measures the SFR on timescales shorter than 10\,Myrs. In the neutral ISM, the photoelectric effect on dust that dominates the heating requires only far UV photons ($\gtrsim 6$\,eV). These photons are also emitted by B-stars and thus [\cii] measures the SFR on much longer timescales. Therefore the phase of the ISM from which the [\cii] line arises from plays a vital role in the relation between the SFR and [\cii].

Various studies have addressed the origins of [\cii] emission. \citet{Croxall2012} presented a pilot study on NGC\,1097 and NGC\,4559 -- two galaxies from the KINGFISH sample \citep[a {\em Herschel} Key program of 61 nearby $\lesssim 30$ Mpc galaxies;][]{Kennicutt2011}. They corrected for the [\cii] emission from the ionized phase by using the [\nii] 205\,\mum\ and 122\,\mum\ lines that arise only from ionized gas (as the ionization potential of nitrogen is 14.53\,eV).  They found that the fraction of [CII] emission coming from photodissociation regions (PDRs) in these galaxies ranges from 0.35 up to 0.8 for regions with warm dust, depending on the assumed electron density.
 Also using the [\nii] lines, \citet{Cormier2012} found for the starburst galaxy Haro\,11 that $\sim$\,40\% of [\cii] emission arises from a diffuse low-ionization gas phase, $\sim$\,20\% from a diffuse neutral phase, and associated the remaining $\sim$\,40\% of emission with PDRs.  
In their observations of M\,51, \citet{Parkin2013} found a similar range of the ionized gas fractions of [\cii], with values of 0.8, 0.7, 0.5 and 0.5 (with a typical uncertainty $\sim$\,0.2), for the nucleus, center, arm and interarm regions of the galaxy, respectively, based on the [\nii] 122\,\mum\ and 205\,\mum\ line observations. Thus observational studies to date suggest $\sim$\,50\% of the [\cii] line may arise from an ionized gas phase, which  agrees with theoretical predictions of 10--50\% from \citet{Abel2006}.

Even if the line arises purely from the neutral ISM, there are further complications in relating [\cii] intensity to SFR.
First, depending on the density and temperature of the gas, cooling via other emission lines can become more efficient than via [\cii] and dominate overall gas cooling \citep[{e.g.~[\oi]\,63\,$\mu$m};][]{Tielens1985}. 
Second, the [\cii] line may suffer from optical depth affects in dense gas \citep{Graf2012}.
Finally, and most importantly, the efficiency with which UV photons heat the gas by the photoelectric heating effect can vary. The variations might occur due to changes in the dust properties, resulting in variable gas heating for a given UV field, or due to change in hardness of the spectrum, resulting in varying heating efficiency for a given dust volume. 
Thus, to properly calibrate the relation between [\cii] and SFR, the origins of the [\cii] line in galaxies must be understood.

Given the observed close relation of the [\cii] line with the UV radiation field, it was also expected that the line and far-IR continuum emission should be related. However, in several galaxies a deficit of the line in relation to the FIR was observed towards high FIR luminosities \citep{Malhotra2001,Helou2001}. This observed FIR line deficit is thought to be associated with a decrease in the efficiency of the photoelectric heating of the gas in FIR luminous objects, caused by changes in the dust grain properties \citep[see][ for a detailed description]{Luhman2003}. 
Based on over 240 luminous infrared galaxies (LIRGs), \citet{DiazSantos2013} found that the [\cii]/FIR ratio decreased with increasing dust temperature. They suggest that this implies that the [\cii] 158\,\mum\ luminosity is not a good indicator of the SFR, as it does not scale linearly with the warm dust emission most likely associated with the youngest stars.
\citet{GraciaCarpio2011} showed that not only [\cii], but also other far-IR lines (e.g. [\oi] 63 and 145\,\mum, [\oiii] 88\,\mum, [\nii] 122\,\mum, from both neutral and ionized phases) exhibit deficits relative to the FIR emission for 44 local starbursts, Seyfert and LIRGs. 

Several theories have been proposed to explain the observed FIR line deficit, for example charged grains reducing the photoelectric effect \citep{Malhotra2001, Croxall2012}, or a reduction in the relative number of polycyclic aromatic hydrocarbons (PAHs) reducing the photoelectric efficiency \citep{Helou2001}, or ionization parameter  \citep{GraciaCarpio2011}. Yet, our understanding of the underlying causes is still incomplete. 
Even after including both correction for diffuse cool neutral and ionized contributions to line emissions and TIR, PDR models used by \citet{Croxall2012} could not reproduce the observed line deficits in the two studied galaxies. \citet{GraciaCarpio2011} discuss the importance of the ionization parameter (higher starlight heating rate U can increase FIR and dust temperature relative to emission of lines), but it cannot alone explain the deficit of all observed lines relative to IR continuum.  
 \citet{Croxall2012} argue that dusty \hii\ regions (elevated dust levels in the ionized gas) are not responsible for line deficits, neither is increased gas density.

In contrast to the standard FIR line deficit, \cite{Israel1996} studied bright \hii\ complexes in LMC and found [\cii]/FIR to be typically around 1\%, considerably higher than found in Galaxy and in most galactic nuclei \citep[i.e. $\sim0.1-1\%$;][and references therein]{Stacey1991}. \cite{Israel1996} explain these higher [\cii]/FIR values as the result of the lower metallicity and lower dust-to-gas ratio in the LMC relative to Galactic regions. Nevertheless, when \cite{Rubin2009} revisited LMC with [\cii] observations by \cite{Mochizuki1994} and FIR by Surveying the Agents of a Galaxy's Evolution \citep[SAGE;][]{Meixner2006, Bernard2008}, they found a flattening of [\cii] as a function of FIR at the FIR high brightness end, a similar trend observed by \citet{Stacey1991}.

One obvious route for understanding the multi-phase origins of [\cii] is to observe the emission at high spatial resolution to resolve the different phases.  This has been the goal of many surveys observing [\cii] within the Milky Way, such as COBE FIRAS \citep{Wright1991,  Bennett1994}, FILM \citep{Makiuti2002} and BICE \citep{Nakagawa1998}. However, these Galactic surveys had to deal with line-of-sight (LOS) confusion along the Galactic disk, making comparisons with stars and other gas tracers difficult.  The recent GOT C+ survey  \citep{Langer2010, Pineda2013, Langer2014} with Herschel-HIFI was able to limit this confusion using their high velocity resolution observations. By comparing their resolved [\cii] data with similarly spectrally resolved CO and HI data they were able, to compare scale heights measured by [\cii], atomic and molecular gas, associate [\cii] emission to the spiral arms between 4 and 11\,kpc, and estimate CO-dark-H$_2$ fraction of the total molecular gas. Yet, while their study is 
extremely powerful with  [\cii] measured in $\sim$\,500 sight-lines in the Galactic plane, the large angular scale of the Milky Way meant that it is impossible to fully map [\cii].

Conversely, in more distant galaxies, multiple ISM phases cannot be separated spatially. Early [\cii] observations of nearby galaxies were typically galaxy averages, comparable to what is now being done with high redshift galaxies \citep{Stacey1991, Malhotra2001, Walter2009, Stacey2010, GraciaCarpio2011}. Recent nearby galaxy studies are only now reaching kiloparsec resolution  \citep{Croxall2012, DiazSantos2013, Parkin2013, Herrera2014}.  

The Local Group represents an ideal compromise between high spatial resolution to study the correlation of [\cii] with various ISM phases, and the simple LOS and galaxy-scale coverage necessary to address these questions. However, most of the Local Group galaxies are low metallicity objects \citep[dwarf galaxies:][]{Israel1996, Kim2002, Rubin2009, Israel2011, Lebouteiller2012}, or low-mass \citep[M\,33, HerM33es project,][]{Kramer2010, Braine2012}, which have significantly different ISM characteristics to massive galaxies in which most of the star formation in the Universe occurs at present.

The Andromeda galaxy (M\,31) provides an ideal target to explore the origins of [\cii] as the only massive, star-forming $L_{\star}$ spiral galaxy in the Local Group. Therefore, as part of a project to understand the heating and cooling of the ISM, we have carried out a \emph{Herschel Space Observatory}\footnote{Herschel is an ESA space observatory with science instruments provided by European-led Principal Investigator consortia and with important participation from NASA.}  far-IR and ground-based optical integral field emission line survey of M\,31 (SLIM; the Survey of lines in M\,31). Previous studies of [\cii] in M\,31 with the \emph{Infrared Space Observatory} targeted the bulge \citep{Mochizuki2000} and a spiral arm on the minor axis \citep{Rodriguez2006}, respectively with far lower effective spatial resolution.

The proximity of M\,31 \citep[$\sim$\,780 kpc,][]{Stanek1998} combined with the \emph{Herschel} resolution of 11\arcsec at 158\,\mum\ (our limiting resolution), enables the study of ISM tracers at sub-kpc scales ($\sim$\,50\,pc), allowing us to spatially separate star forming from diffuse regions. The large amount of available ancillary data, and the simplicity provided by an external perspective, make M\,31 a unique target for understanding [\cii] emission. Our survey targeted several different star forming regions across the disk of M\,31, enabling a study of the relative variation of [\cii] emission (and heating and cooling of the ISM in general) over a wide scope of physical conditions such as stellar surface densities, SFRs, and metallicities. The only caveat is that due to high inclination of M\,31 \citep[$70^{\circ}$;][]{Dalcanton2012}, our data also suffer from some LOS confusion. Nevertheless, our analysis is valid within the known limitations, that [\cii] emission line: 1) does not spatially 
resolve individual \hii\ regions, nor PDRs, 2) is not spectrally resolved ($\sim200$km/s.)

The paper is organized as follows: the [\cii] and ancillary data acquisition is described in \S~\ref{sec:data}, along with further processing details. In \S~\ref{sec:result_cii_sfreg}, we present the spatial decomposition of several ISM and SFR tracers. In \S~\ref{sec:result_cii_sfr} we test the calibration of [\cii] to SFR relation at high spatial resolution, and we compare to existing studies on different spatial scales. In \S~\ref{sec:fir_def}, we investigate the FIR line deficit and its relation to the ability of [\cii] to track SFR. Finally, we discuss our results in \S~\ref{sec:disc}, and present our conclusions in \S~\ref{sec:concl}.

\section{Data}
\label{sec:data}

M\,31 is the most massive external galaxy in the Local Group.
It is a highly inclined \citep[i.e., 70$^{\circ}$;][and references therein]{Dalcanton2012} spiral galaxy classified as SA(s)b (see Tab.~\ref{tab:m31}) and presents ring-like structure. Due to its proximity \citep[$\sim$\,780\,kpc;][]{Stanek1998}, it is possible to reach a high spatial resolution with \emph{Herschel} ($\sim$\,50\,pc at 160\,\mum).

\begin{deluxetable}{cc}
\tablewidth{0.3\textwidth}
\tablecolumns{2}
\tablecaption{M\,31 information}
\tablehead{ }
\startdata
Nucleus position\tablenotemark{a}   & $00^h42^m44.^s35$                  \\ 
(J2000)                             & $+ 41^{\circ}16\arcmin08\farcs60$  \\
Inclination\tablenotemark{b}        & 70$^{\circ}$                       \\
Position angle\tablenotemark{b}     & 43.2$^{\circ}$                     \\
Distance\tablenotemark{c}           & $780 \pm 40$\,kpc                  \\
Morphological type\tablenotemark{a} & SA(s)b                             \\
SFR\tablenotemark{d}                & 0.4\,M$_{\odot}$/yr  
\enddata
\label{tab:m31}
\tablenotetext{a}{Based on NED data and references therein}
\tablenotetext{b}{\citet{Dalcanton2012}}
\tablenotetext{c}{\citet{Stanek1998}}
\tablenotetext{d}{\citet{Barmby}}
\end{deluxetable}

The measured SFR in M\,31 is low, $\sim$\,0.4\,M$_{\odot}$ yr$^{-1}$, over the last 100\,Myr \citep{Barmby}, and is concentrated mostly in the spiral arms and rings. In addition, a large diffuse fraction of H$\alpha$ emission is also observed \citep{Walterbos1994,Azimlu2011}.

We carried out a far-IR and optical survey of interstellar emission lines in M\,31 (SLIM; the Survey of lines in M\,31; PI K. Sandstrom) to study the cooling emission from a variety of ISM phases. 
The survey consists of five 3\arcmin$\times$3$\arcmin$ ($\sim$700$\times$700\,pc) Fields with Herschel PACS spectroscopy and optical integral field spectroscopy.
This line survey is complemented with infrared photometry from {\em Herschel} and {\em Spitzer}. We describe the data we used in our present analysis in the following subsections.

\subsection{{\em Herschel} PACS spectroscopy}
\label{sec:data_pacs}

\begin{figure*}[!htp]
\begin{center}
\includegraphics[width=1.\textwidth]{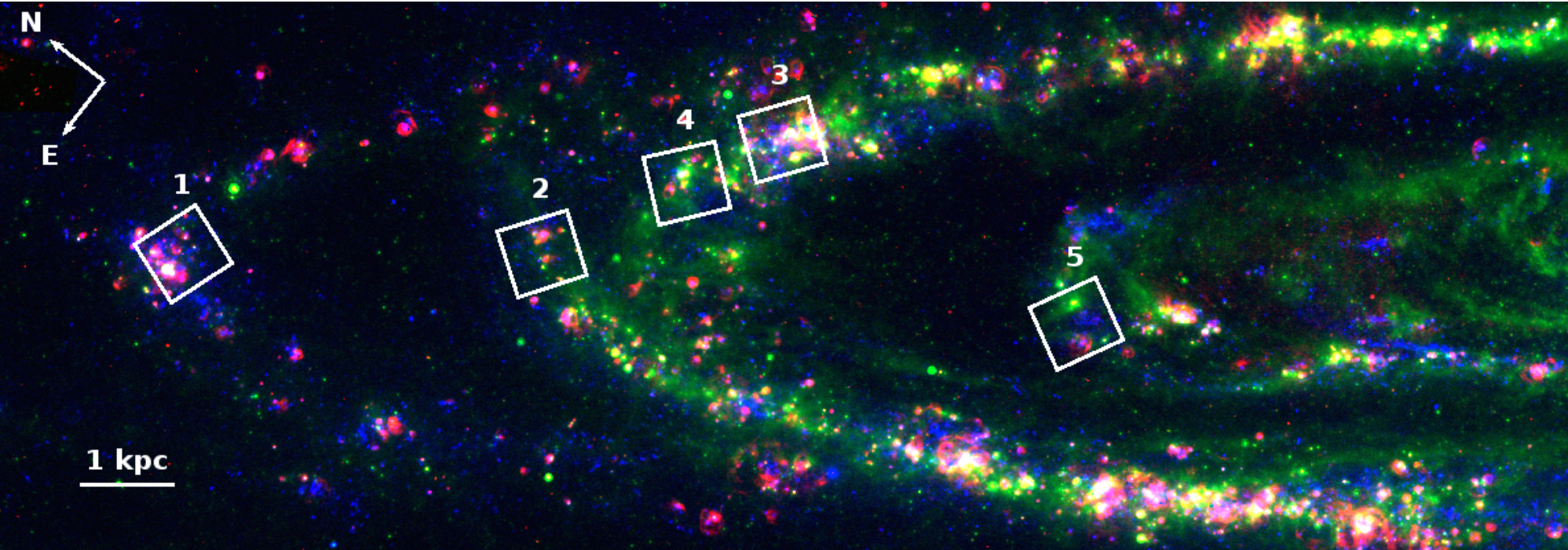}
\caption{Position and orientation of the five Fields targeted in this study overlaid on a RGB image (blue - Galex FUV, green - MIPS 24\,\mum\ and red - H$\alpha$), region numbers are labeled and the orientation is shown in the upper left. A scale bar indicates 1\,kpc (4.4\arcmin).}
\label{fig:coverage}
\end{center}
\end{figure*}

\begin{deluxetable}{ccccc}
\tablewidth{0pt}
\tablecolumns{5}
\tablecaption{Coordinates of Field centers}
\tablehead{\multicolumn{1}{c}{F} & 
\multicolumn{1}{c}{R.A.} & 
\multicolumn{1}{c}{Dec.} & 
\multicolumn{1}{c}{P.A.} & 
\multicolumn{1}{c}{Observation\tablenotemark{a}}  \\
\multicolumn{1}{c}{} & 
\multicolumn{1}{c}{(J2000)} & 
\multicolumn{1}{c}{(J2000)} &
\multicolumn{1}{c}{[$^{\circ}$]} &
\multicolumn{1}{c}{ID} }
\startdata
1     & $00^h46^m29.^s17$ & $+42^{\circ}11\arcmin30\farcs89$ & 70.7  & 1342236285 \\
2     & $00^h45^m34.^s79$ & $+41^{\circ}58\arcmin28\farcs54$ & 55.7  & 1342238390  \\
3     & $00^h44^m36.^s49$ & $+41^{\circ}52\arcmin54\farcs21$ & 55.0  & 1342238391  \\
4     & $00^h44^m59.^s26$ & $+41^{\circ}55\arcmin10\farcs47$ & 51.0  & 1342238726 \\
5     & $00^h44^m28.^s76$ & $+41^{\circ}36\arcmin58\farcs91$ & 63.0  & 1342237597 
\enddata
\label{tab:obs}
\tablenotetext{a}{[\cii] \& [\oi] observations acquired between 28$^{\rm th}$ Feb and 1$^{\rm st}$ Mar, 2012}
\end{deluxetable}

Observations of the [\cii] 158\,\mum, [\oi] 63\,\mum, and [\nii] 122\,\mum\ spectral lines were carried out using the unchopped mapping mode on the Photodetector Array Camera and Spectrometer \citep[PACS;][]{Poglitsch2010} on board of the ESA {\em Herschel} Space Observatory \citep{Pilbratt2010} for a total of 47.1 hours of observing time (OT1\_ksandstr\_1; PI K. Sandstrom). Due to the large angular extent of M\,31, galaxy emission fell within all available chopper-throws. Therefore we used the unchopped mode with an off position defined to be well outside of the body of M\,31, and visited 3 times during each AOR. 
The [\cii] 158\,\mum\ and [\oi] 63\,\mum\ lines were mapped in five Fields of $3\arcmin$ by $3\arcmin$ (700\,pc $\times$ 700\,pc) using rastering in the instrument reference frame with 37\farcs5 and 23\farcs5 steps. We choose $3\arcmin$ by $3\arcmin$ Field sizes: (1) to cover the scale over which energy from star-forming regions should be deposited in the ISM and (2) to match the approximate size of a resolution element in the KINGFISH [\cii] maps at the sample's average distance of 12\,Mpc. 

The five Fields probe different physical conditions along M\,31's major axis, sampling different H$\alpha$, FUV, and 24\,\mum\ surface brightnesses and atomic to molecular gas ratios, while still focusing on regions of active star-formation. The Fields are shown in Figure~\ref{fig:coverage} and are tabulated in Table~\ref{tab:obs}. 
All Fields lie on the NE major axis of the galaxy as this side is covered by the Pan-chromatic Hubble Andromeda Treasury program  \citep[PHAT;][]{Dalcanton2012}, providing a large, high-resolution, UV to NIR ancillary dataset. We label these Fields F1 (outermost) through F5 (innermost), though we note that this order is not entirely radial as F3 is at a slightly smaller galactocentric radius than F4. F1 covers a star-forming region in the outer spiral arm at $\sim$\,16\,kpc, while Fields F2 to F4 fall in the star forming ring at $\sim$\,10\,kpc, and F5 covers a region in the inner arm at $\sim$\,7\,kpc, from the center of the galaxy, respectively. 

The [\nii] 122\,\mum\ line was observed only in six smaller maps of $1\arcmin \times 1\arcmin$ that targeted the brightest \hii\ regions in the [\cii] Fields (2 in F1, 1 in F2, 2 in F3 and 1 in F4). Unfortunately, the [\nii] line was found to be very weak and was detected only in very few spaxels.

Similarly, though the [\oi] line was significantly detected in several regions in all Fields, it was found to be weaker than [\cii] in all reliably detected regions, with typical values of [\oi]/[\cii]\,$\sim 0.46$. We find these reliable detections only in the brightest regions (SF). We expect [\oi]/[\cii] to be even less in the diffuse regions \citep[see Figure~9.2 in][]{Tielens2005}. In the following, we focus the analysis mostly on the [\cii] 158\,\mum\ line and leave a discussion of the [\oi] and [\nii] lines for a future paper (Kapala et al. 2015 in prep). 

\begin{figure*}[ht!]
\begin{center}
\includegraphics[width=.9\textwidth]{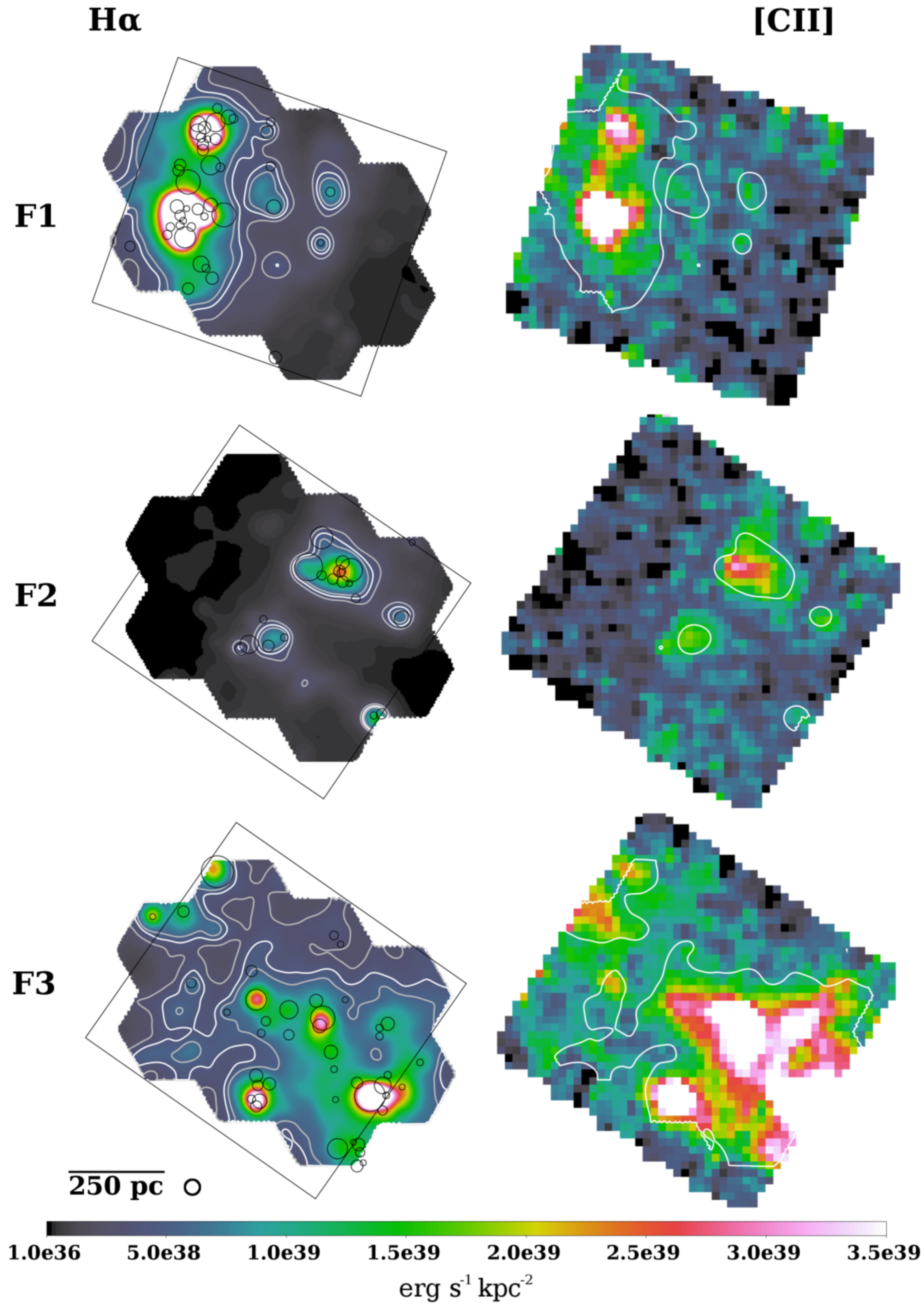}
\caption{{\bf Left column:} H$\alpha$ emission from SF regions in Fields 1 to 3,  convolved to match [\cii] resolution. Black circles indicate the \hii\ regions defined in the \citet{Azimlu2011} catalog, with the radius of the circles set to the FWHM of the \hii\ regions. White contour level indicates chosen level L$_0$ that delineates SF regions, two grey ones are ${\rm L_0 \pm 30\% \, L_0}$. Black boxes show [\cii] pointings. The [\cii] beam size (11.0\arcsec) and scale bar (66.1\arcsec) are indicated in the bottom left corner. {\bf Right column:} {\em Herschel} PACS [\cii] 158\,\mum\ maps of Fields 1 to 3 with L$_0$ H$\alpha$ contour overlaid. All maps have a common linear color scale given at the bottom of the figure. Note that only pixels present in both [\cii] and H$\alpha$ maps are used in our analysis.}
\label{fig:ha_cii_maps}
\end{center}
\end{figure*}

\begin{figure*}[ht!]
\begin{center}
\includegraphics[width=.9\textwidth]{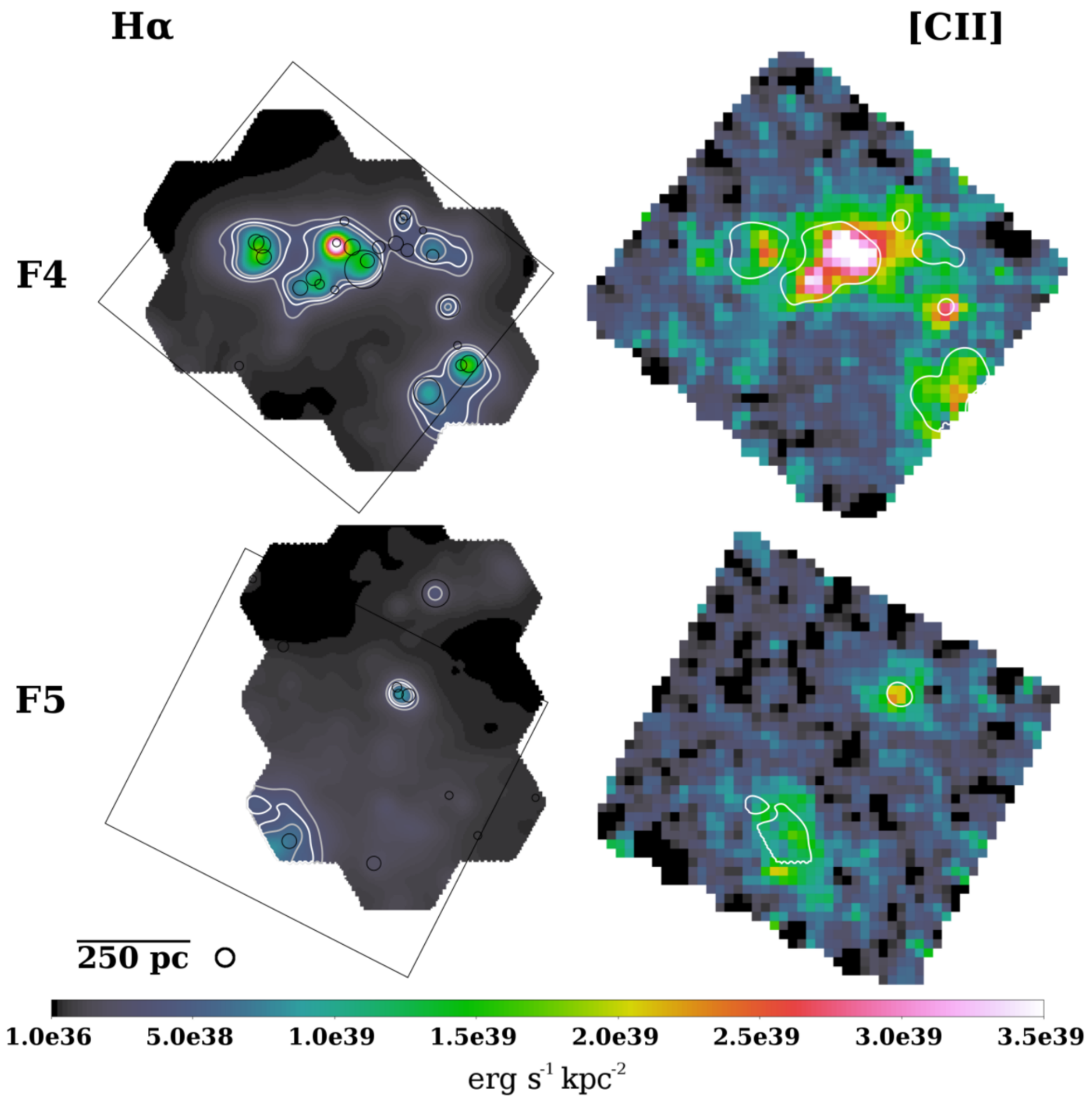}
\caption{Same as Figure~\ref{fig:ha_cii_maps} for Fields 4 and 5.}
\label{fig:ha_cii_maps2}
\end{center}
\end{figure*}

 The [\cii] data was reduced using the {\em Herschel} Interactive Processing Environment (HIPE) version 8.0 \citep{Ott2010}. Reductions applied the standard spectral response functions, flat field corrections, and flagged instrument artifacts and bad pixels \citep[see ][]{Poglitsch2010}. The dark current, determined from each individual observation, was subtracted during processing because it was not removed via chopping. Transient removal was performed using the surrounding continuum, as described in \citet{Croxall2012}. 
In-flight flux calibrations were applied to the data. After drizzling in the HIPE pipeline, the [\cii] 158\mum\ (FWHM\,$\sim$\,11\farcs0) line was integrated in velocity to produce maps with 2\farcs6 pixels.  For the analysis, we rebinned the maps to have approximately half-beam spaced pixel size (5\farcs2).  The final integrated intensity maps of the [\cii] 158\,\mum\ emission line for each Field are shown in the right side in Figures~\ref{fig:ha_cii_maps} and~\ref{fig:ha_cii_maps2}.

The details of the observations; coordinates, AOR numbers, are listed in Table~\ref{tab:obs}.  The mean 1-$\sigma$ [\cii] surface brightness sensitivity of the line integrated intensity in all pixels in the overlapping regions of the [\cii] and H$\alpha$ Fields is $1.46 \times 10^{38} \ {\rm erg \ s^{-1} kpc^{-2}}$ with standard deviation $5.51 \times 10^{37} \ {\rm erg \ s^{-1} kpc^{-2}}$. Note that individual points might have values below that average limit. That is because for an application of the significance cuts in \S~\ref{sec:result}, we use each pixel's 1-$\sigma$ noise measured from its spectrum, which accounts for the goodness of the line fit and PACS scanning flaws, not only instrumental sensitivity limit. The absolute [\cii] flux calibration uncertainty is $\sim$\,30\%\footnote{http://herschel.esac.esa.int/twiki/pub/Public/PacsCalibration\\Web/PacsSpectroscopyPerformanceAndCalibration\_v2\_4.pdf}.
We visually inspected the spectral cubes in low S/N regions and find that the quoted uncertainties on the line flux are reasonable.

\subsection{PPAK IFS}
\label{sec:data_ppak}

We obtained optical integral field spectroscopy (IFS) covering the same five Fields as the PACS spectral maps (PI K. Sandstrom), over nine nights in September 2011 using the Calar Alto 3.5m telescope with the PMAS instrument in PPAK mode with the V300 grating \citep{Roth2005, Kelz2006}.  This setup provides 331 science fibers, each 2\farcs68 in diameter, that sample a spectral range from 3700--7000\,\AA\  with $\sim$\,200 km s$^{-1}$ instrumental resolution and hexagonally tile a $\sim$\,1\arcmin field of view.  Our observation and reduction procedures follow very closely those outlined in \citet{Kreckel2013}, and we summarize here only the key steps and variations from that description.

Each of the five Fields were mosaicked with 10 PPAK pointings.  To completely recover the flux and fill in gaps between the fibers, each pointing was observed in three dither positions.  We reduced these nearly 50,000 spectra using the {\tt p3d} software package \citep{Sandin2010}.  All frames are bias subtracted, flat-field corrected and wavelength calibrated using standard calibration observations.  Frames are cleaned of cosmic rays using the L. A Cosmic technique \citep{vanDokkum2001} as adapted within {\tt p3d}.   Spectra are extracted using a modified optimal extraction method that simultaneously fits all line profiles with a Gaussian function \citep{Horne1986}.  Relative flux calibration is applied using one of two standard stars observed during the night, where we choose the star that appears most centered within a single fiber.  As M\,31 is quite extended on the sky, separate sky pointings were obtained and a best fit linear combination from the set of sky pointings observed that night are used to 
optimally subtract the sky emission from the science spectra.     

Seeing was sub-fiber ($<$3\arcsec) and astrometry for each mosaic position has been applied through comparison by eye of features in our H$\alpha$ maps with Local Group Galaxies Survey H$\alpha$ images \citep{Azimlu2011}.  From comparison of stellar sources within the PPAK data and SDSS \citep{Aihara2011} broadband images we estimate our astrometry is accurate to within 1\arcsec, sufficient for comparison with the lower resolution PACS images.
All observations were taken at airmass below 2, with nearly all below 1.5, so we neglect the effects of differential atmospheric refraction.   Conditions were not consistently photometric, so we have re-calibrated the flux scaling of each dither position and scaled it to the continuum subtracted Local Group Galaxies Survey H$\alpha$ images \citep{Azimlu2011}.  We expect our relative flux calibration to be accurate to within 5\%, however, we allow for a larger systematic uncertainty of 20\%.

\begin{figure*}[!ht]
\begin{center}
\includegraphics[width=1.\textwidth]{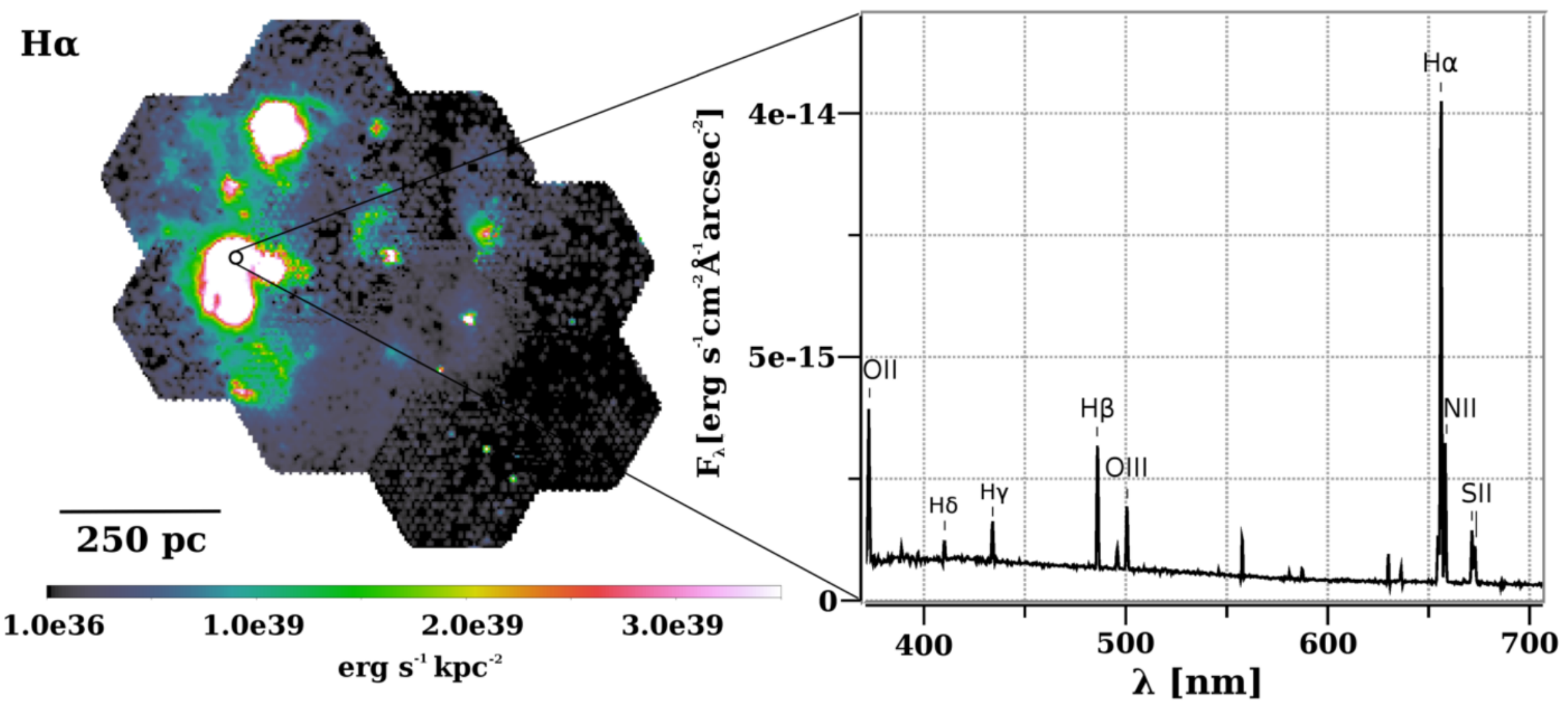}
\caption{PPAK IFS data in Field 1. Left panel - H$\alpha$ map at the native resolution, right panel - spectrum in the pixel marked as black circle (RA(J2000) $= 00^h46^m34.^s52$, Dec.(J2000) $= +42^{\circ}11\arcmin43\farcs82$).}
\label{fig:ha_spec}
\end{center}
\end{figure*}

The H$\alpha$ line was measured in each spectrum using the GANDALF package \citep{Sarzi2006}, which employs penalized pixel fitting \citep[{\tt pPXF};][]{Cappellari2004} to simultaneously fit both  stellar continuum templates and gaussian emission line profiles.  This allows us to deblend the contribution of [\nii] from the H$\alpha$ emission line, as well as to correct for underlying stellar absorption.  We use here simple stellar population (SSP) template spectra from the \citet{Tremonti2004} library of \citet{Bruzual2003} templates for a range of 10 ages (5 Myr - 11 Gyr) and two metallicities (1/5 and 1 solar), though using stellar templates selected from the MILES library \citep{MILES2011} does not significantly change our measured line fluxes.  These  H$\alpha$ line fluxes are interpolated onto a regular grid using Delaunay linear triangulation, resulting in images with $\sim$\,2\farcs5 resolution that reach 3$\sigma$ H$\alpha$ surface brightness sensitivities of ${\rm 2 \times 10^{-16}\ erg\ s^{-1} cm^{
-2} arcsec^{-2}}$.  As in the case of [\cii] data, one of the product of the reduction pipeline is noise map. The H$\alpha$ noise depends not only on the line strength, but also on the observing conditions. 
Due to atmospheric variations, we compare the values between the individual pointings. This approach returns median noise ${\rm 3.60\times10^{37} \ erg \ s^{-1} kpc^{-2}}$ and standard deviation ${\rm 3.15\times10^{37}\  erg \ s^{-1} kpc^{-2}}$. 
Figure~\ref{fig:ha_spec} shows the map for Field 1, as well as an extracted spectrum.

\subsection{{\em Spitzer} \& {\em Herschel} Photometry}
\label{sec:data_ir}

We make use of the data from several infrared surveys of M\,31 using {\em Herschel} and {\em Spitzer}.  M\,31 was observed with the Multiband Imaging Photometer \citep[MIPS;][]{Rieke2004} on board of the {\em Spitzer} Space Telescope \citep{Werner2004} by \citet{Gordon2006} and the InfraRed Array Camera \citep[IRAC;][]{Fazio2004} by \citet{Barmby}.  In addition, M\,31 was observed by the {\em Herschel} Space Observatory (PI O. Krause) using the PACS and SPIRE instruments \citep[Spectral and Photometric Imaging Receiver]{Griffin2010}.  For details of the PACS and SPIRE observations and processing see \citet{Groves2012} or \citet{Draine2014}.

PACS images in all photometric bands (70, 100 and 160\,\mum) were obtained in slow parallel mode, with a final image size of $\sim$\,3$^{\circ} \times 1^{\circ}$ aligned with the position angle of M\,31's major axis. This covers M\,31's full disk including the 16\,kpc ring.  All images were reduced to level one using HIPE v6.0, and then SCANAMORPHOS v12.0 \citep{Roussel2012} was used to produce the final product.

The FWHM is 5\farcs6 for 70\,\mum, 6\farcs8 for PACS 100\,\mum\ and 11\farcs4 for PACS 160\,\mum. Beam sizes are taken from PACS Observer's Manual\footnote{${\rm http://herschel.esac.esa.int/Docs/PACS/html/pacs\_om.html}$} for 20\arcsec s$^{-1}$ scans. 1\arcsec pixel size was used for all PACS bands.  

For the {\emph Spitzer} bands we considered, the IRAC 8\,\mum\ has a FWHM=2\farcs0 and the MIPS 24\,\mum\ has a FWHM=6\farcs5.

To remove the foreground emission from the Milky Way and any other foregrounds or backgrounds, we measured the median surface brightness in regions on the edges of the map, away from the main body of M\,31.  These values showed no clear gradient, so a uniform background was subtracted from each image. The determined backgrounds were: 2.40 MJy ${\rm sr}^{-1} (8\,\mum)$, -0.0043 MJy ${\rm sr}^{-1} (24\,\mum)$, 3.17 MJy ${\rm sr}^{-1} (70\,\mum)$, 3.23 MJy ${\rm sr}^{-1} (100\,\mum)$ and 2.29 MJy ${\rm sr}^{-1} (160\,\mum)$.

\subsection{Further processing}
\label{sec:data_proc}

All maps were convolved to match the PACS 160\,\mum\ resolution of 11\farcs0, using the \citet{Astropy2013} package {\tt convolve\_fft} and kernels from \citet{Aniano2011} for each specific filter: IRAC 8\,\mum, MIPS 24\,\mum, PACS 70\,\mum, PACS 100\,\mum, and PACS 160\,\mum. The latter kernel was also used for the [\cii] map. For the H$\alpha$ images we assumed an intrinsic Gaussian PSF with a FWHM of 2\farcs5 \citep[see][]{Kreckel2013}, and used the corresponding convolution kernel. The convolved maps were all resampled to match the [\cii] pixel size of 5\farcs2 using the {\tt Montage} \citep{Jacob2010} Python wrapper\footnote{http://www.astropy.org/montage-wrapper}. For any direct comparison, the units of the images were converted to erg s$^{-1}$ cm$^{-2}$ sr$^{-1}$ or erg s$^{-1}$ kpc$^{-2}$.

\subsection{Derived Quantities}
\label{sec:quan}

We use the \citet{Calzetti2007} formula to calculate SFR surface density from a linear combination of H$\alpha$ and 24\,\mum\ emission, which includes a correction for absorbed emission that is important when young stars are embedded in their natal clouds. At the physical resolution we reached in this work, we begin to resolve out the \hii\ regions and diffuse gas, such that any SFR calibration based on linear combination of star-forming tracers no longer strictly holds \citep{Leroy2012, Simones2014}.  Furthermore, SFR measurements at 50\,pc scales are problematic using indirect tracers like H$\alpha$ and dust emission for many reasons, including sampling of the stellar population \citep[others include drift of stars between pixels, star formation history, and others outlined in ][]{KennicuttEvans}.
For the purposes of our work, we are primarily interested in the relative spatial distribution of these SFR tracers and not an accurate measurement of the 50\,pc scale SFR.

\begin{equation}
\begin{multlined}
\rm{\Sigma_{SFR}} \ \rm{[ M_{\odot} yr^{-1} kpc^{-2}] =} \\ 
( 634 \ I_{\rm H\alpha } + 19.65 \ I_{\rm 24\mu m} ) \ \rm{[ erg \ s^{-1} cm^{-2}sr^{-1}]} 
\end{multlined}
\label{eq:sfr}
\end{equation} 
This calibration assumes a constant SFR for 100\,Myr, solar metallicity, and the stellar population models and Kroupa initial mass function (IMF) from Starburst99 \citep[2005 update;][]{Leitherer1999}.  In reality, the SFR in M\,31 has not been constant over the last 100\,Myr, with the star formation histories based on Hubble stellar photometry showing significant variation in this timescale, and across the disk in M\,31 \citep{Lewis2014}. Nevertheless, as H$\alpha$ is the dominant contributor to the SFR shown for most of the points in Figure~\ref{fig:ciisfr50}, therefore the effect of the variation in the star formation history to this calibration is limited. The average SFR surface densities, listed in Table~\ref{tab:sfr}, range between $2-7\times10^{-3}$ M$_{\odot}$ yr$^{-1}$ kpc$^{-2}$, the estimated median uncertainty is $4.2\times10^{-4}$ M$_{\odot}$ yr$^{-1}$ kpc$^{-2}$ with the standard deviation $3.3\times10^{-4}$ M$_{\odot}$ yr$^{-1}$ kpc$^{-2}$.

\begin{deluxetable}{cccccc}
\tablewidth{0pt}
\tablecolumns{6}
\tablecaption{The average [\cii] emission, $\Sigma_{\rm SFR}$ and metallicity of the Fields}
\tablehead{\multicolumn{1}{c}{Field} & 
\multicolumn{1}{c}{R\tablenotemark{a}} & 
\multicolumn{1}{c}{R} & 
\multicolumn{1}{c}{[\cii]$_{\rm TOT}$\tablenotemark{b}} &
\multicolumn{1}{c}{${\rm \Sigma_{SFR}}$\tablenotemark{c}} & 
\multicolumn{1}{c}{[O/H]\tablenotemark{d}} \\
\multicolumn{1}{c}{} & 
\multicolumn{1}{c}{[kpc]} & 
\multicolumn{1}{c}{[$^{\circ}$]} &
\multicolumn{1}{c}{} &
\multicolumn{1}{c}{} &
\multicolumn{1}{c}{} }
\startdata
1     & 16.03  & 1.177   & 8.26      &  4.740           & 0.774 \\
2     & 12.28  & 0.902   & 5.81      &  1.735           & 0.943 \\
3     & 11.31  & 0.831   & 17.01     &  7.136           & 0.993 \\
4     & 11.45  & 0.841   & 7.93      &  2.916           & 0.985 \\
5     & 6.86   & 0.504   & 4.17      &  1.944           & 1.361 
\enddata
\label{tab:sfr}
\tablenotetext{a}{Galactocentric radius}
\tablenotetext{b}{Average Field [\cii] surf. brightness in ${\rm [ 10^{38} \ erg\ s^{-1} kpc^{-2}] }$}
\tablenotetext{c}{Average Field SFR surface density in $[10^{-3} {\rm  \ M_{\odot}yr^{-1}kpc^{-2}}]$}
\tablenotetext{d}{Metallicity relative to solar with ${\rm \log(O/H)_{\odot}=-3.31}$}
\end{deluxetable}

The average gas phase metallicity for each Field is calculated from Equation 7 from \citet{Draine2014} and listed below, using the deprojected radial distance of the Field center. The \citet{Draine2014} formula is equivalent to the ``direct T$_e$-based method'' to derive metallicity gradient measured by \citet{Zurita2012} in range of radii where both have measurements. \citet{Zurita2012} use auroral optical line' ratios in \hii\ regions (for $12+\log(O/H)_{\odot}=8.69$), while \citet{Draine2014} use the dust-to-gas ratio as a metallicity proxy:
\begin{equation}
\frac{(O/H)}{(O/H)_{\odot} } \approx  \left\{
\begin{array}{l l}
1.8  \exp(-R/19\text{kpc})   & \text{for } R<8 \text{kpc}    \\
3.08 \exp(-R/8.4\text{kpc}) & \text{for } 8<R<18 \text{kpc.} \nonumber
 \end{array} \right.\
\end{equation}
The metallicity ranges from 0.8 to 1.4 Z$_{\odot}$, and the individual results per Field are listed in Table~\ref{tab:sfr}.

We calculate the total infrared emission (TIR) using a linear combination of IRAC 8 and 24\,\mum, PACS 70 and 160\,\mum\   (all in units of $\rm{erg \ s^{-1} cm^{-2}sr^{-1}}$), as described in \citet{DraineLi2007}:  
\begin{equation}
TIR = 0.95 (\nu I_{\nu})_8 + 1.15(\nu I_{\nu})_{24} + (\nu I_{\nu})_{70} + (\nu I_{\nu})_{160}, \nonumber
\end{equation}
where the worst-case error in estimating TIR is $\sim$\,30\%, and dominated by noise and calibration uncertainties of its components.  This TIR calibration is identical to what was assumed by \citet{Croxall2012}, whose results we compare in \S~\ref{sec:fir_def700}.

\section{Results}
\label{sec:result}

\subsection{[\cii] from SF Regions}
\label{sec:result_cii_sfreg}

As discussed above, the [\cii] emission comes from multiple phases of the ISM, which includes \hii\ regions and their bordering PDRs, as well as the diffuse ISM (cold neutral medium -- CNM, warm neutral medium -- WNM, and warm ionized medium -- WIM). Our [\cii] maps have a resolution of $\sim$\,50\,pc. Individual \hii\ regions in our Fields have typical sizes of $\sim$\,20--30 pc in diameter \citep{Azimlu2011} and PDRs give an additional $\sim$\,0.3--3\,pc layer, assuming carbon is ionized up to A$_V \sim 5$ \citep{Tielens2005} into the PDR for a typical density $n_H \sim 10^3-10^4 \ {\rm cm^{-3}}$.  Therefore, \hii\ regions and PDRs are blended together at our working resolution. Hereafter, we refer to these ``blended'' \hii\ regions and PDRs as star-forming (SF) regions. Although we cannot separate \hii\ regions and PDRs in this study, we can investigate the fraction of [\cii] arising from the diffuse ISM versus SF regions. In the following Section, we describe how we delineate SF regions using our H$\
alpha$ maps.

\subsubsection{Delineating SF Regions}
\label{sec:result_reg}

\hii\ regions in M\,31 can be identified and resolved \citep{Azimlu2011} using H$\alpha$ emission. Because of the much lower resolution of the [\cii] map, multiple \hii\ regions might reside in a single pixel.  Therefore, we use the \hii\ region catalogs as a guide to define contours in H$\alpha$ emission, at a resolution matched to the [\cii] maps, that enclose most of the massive star formation.
H$\alpha$ emission was previously used to distinguish between \hii\ regions and diffuse media for distinguishing the origin of [\cii] in LMC by both \citet{Kim2002} \citep[following work of ][]{Kennicutt1986} with H$\alpha$ contour $5.09\times10^{39} \ {\rm \ erg\ s^{-1} kpc^{-2} }$, and by \citet{Rubin2009} with $1.20\times10^{39} \ {\rm \ erg\ s^{-1} kpc^{-2} }$.

H$\alpha$ emission from \hii\ regions in M\,31 is relatively modest, compared to other Local Group galaxies. \citet{Azimlu2011} report a total H$\alpha$ luminosity coming from \hii\ regions in M\,31 $\sim$\,$1.77 \times 10^{40}$ erg s$^{-1}$, which is comparable to the luminosity of the 30 Doradus complex in the LMC alone $\sim$\,$1.5 \times 10^{40}$ erg s$^{-1}$ \citep{Kennicutt1984}. 
On average, \citet{Azimlu2011} deduced a $\sim$\,63\% diffuse ionized gas contribution from spatially separating \hii\ regions from the surrounding emission.  \citet{Walterbos1994} used [\sii]/H$\alpha$ ratios to identify diffuse ionized gas and obtained a $\sim$\,40\% contribution.

To start, we convolve our PPAK H$\alpha$ maps to the [\cii] resolution. Then, we use the \hii\ region catalog from the Local Group Survey \citep[LGS;][]{Azimlu2011} to identify the location and half-light radii of the \hii\ regions.  We use the LGS catalog rather than our PPAK maps because of the 2 times higher spatial resolution of their imaging, which allows for better identification of \hii\ regions. 
Nevertheless, for the remaining analysis, we use our PPAK H$\alpha$ maps rather than the narrowband imaging from LGS for the following reasons: (1) [\nii] 6548 and 6583\,\AA \ are blended with H$\alpha$ in the narrow band imaging (see Figure~\ref{fig:ha_spec}) and variations in the H$\alpha$/[\nii] ratio make correcting for blending difficult and (2) removal of stellar continuum is significantly easier from the PPAK data compared to the narrowband imaging.

We then define a surface brightness threshold in the convolved H$\alpha$ map above which  the majority of the \hii\ regions identified by \citet{Azimlu2011} are enclosed.
We display the contour at the selected surface brightness threshold, $L_0= 4.19\times 10^{38}{\rm \ erg\ s^{-1} kpc^{-2} }$, in Figures \ref{fig:ha_cii_maps} and \ref{fig:ha_cii_maps2}. Our H$\alpha$ threshold is lower than the values used by \citet{Kennicutt1986,Kim2002} and \citet{Rubin2009}, as we wished to definitively encompass both the \hii\ regions and their associated PDRS in our ``SF regions''. Table~\ref{tab:sfregions} lists the fraction of \hii\ regions enclosed by this contour, along with the areal fraction enclosed, for each Field. In Field 5 several very faint \hii\ regions lie outside the contour, but these regions make a negligible contribution to the overall H$\alpha$ emission. Finally, we use this contour and calculate the fractions of the pixel area within the boundary. ``Diffuse'' regions are defined as lying outside the H$\alpha$ contour.

\begin{deluxetable*}{ccccccc}
\tablewidth{0.68\textwidth}
\tablecolumns{7}
\tablecaption{Ratios of fraction of emission of the ISM tracers coming from SF regions over total emission per field, for a H$\alpha$ contour $L_0= 4.19\times 10^{38}{\rm \ erg\ s^{-1} kpc^{-2} }$  defining SF region}
\tablehead{\multicolumn{1}{c}{Field} & 
\multicolumn{1}{c}{$\rm{\frac{H\alpha_{SF}}{H\alpha_{TOT}}}$} & 
\multicolumn{1}{c}{$\rm{\frac{[CII]_{SF}}{[CII]_{TOT}}}$ } & 
\multicolumn{1}{c}{$\rm{\frac{M24_{SF}\tablenotemark{a}}{M24_{TOT}}}$} &
\multicolumn{1}{c}{$\rm{\frac{TIR_{SF}}{TIR_{TOT}}}$} & 
\multicolumn{1}{c}{$\frac{A_{SF}\tablenotemark{b}}{A_{TOT}}$} &
\multicolumn{1}{c}{$\frac{N_{HII}(SF)\tablenotemark{c}}{N_{TOT}(F)}$} \\
\multicolumn{1}{c}{} & 
\multicolumn{1}{c}{[\%]} & 
\multicolumn{1}{c}{[\%]} &
\multicolumn{1}{c}{[\%]} &
\multicolumn{1}{c}{[\%]} &
\multicolumn{1}{c}{[\%]} &
\multicolumn{1}{c}{} }
\startdata
1  &  82.57  &  63.04  &  70.31  &  34.72  &  64.41 & 40/41 \\
2  &  44.52  &  20.15  &  14.40  &  7.04   &  13.18 & 17/20 \\
3  &  82.35  &  80.49  &  79.63  &  60.02  &  77.53 & 39/41 \\
4  &  56.23  &  30.97  &  26.40  &  13.47  &  23.07 & 20/25 \\
5  &  21.90  &  12.72  &  7.85   &  4.33   &  6.67  & 4/11
\enddata
\label{tab:sfregions}
\tablenotetext{a}{MIPS 24\,\mum}
\tablenotetext{b}{SF region areal fraction of Field}
\tablenotetext{c}{The number of \hii\ regions enclosed by SF regions to the total number of \hii\ regions in the Field}
\end{deluxetable*}

We apply the above procedure to calculate the fraction of our tracers (H$\alpha$, [\cii], 24\,\mum\, TIR) spatially associated with SF regions. To estimate the uncertainties on this fraction, we moved the contour level by $\pm30$\% of the chosen H$\alpha$ surface brightness and recalculate the fractions.  We find that $\pm30$\% defines a reasonable range of potential contours surrounding the \hii\ regions as can be seen in Figure~\ref{fig:ha_cii_maps}.  The uncertainty in defining the contour is the dominant component of the total uncertainty on the diffuse fractions since calibration uncertainties divide out and the S/N is high. 

We note that the majority of the following analysis compares the relative concentration of H$\alpha$, [\cii], 24\,\mum\ and TIR emission in SF regions, rather than focusing on the absolute value of the fraction. For this reason, our results are not sensitive to the exact definition of the contour level, as it mostly represents a fiducial level for comparing the extent of the various SFR tracers.
We note that the absolute values of the fractions are sensitive as well to the size of the maps, as diffuse emission might extend beyond the limits of the maps. However, this issue do not detract from our results, because our Field sizes are representative of individual resolution elements in nearby galaxy surveys, therefore can be directly compared.

\subsubsection{Fraction of [\cii] and other tracers from SF Regions}
\label{sec:result_frac}

\begin{figure}[!ht]
\begin{center}
\includegraphics[width=.5\textwidth]{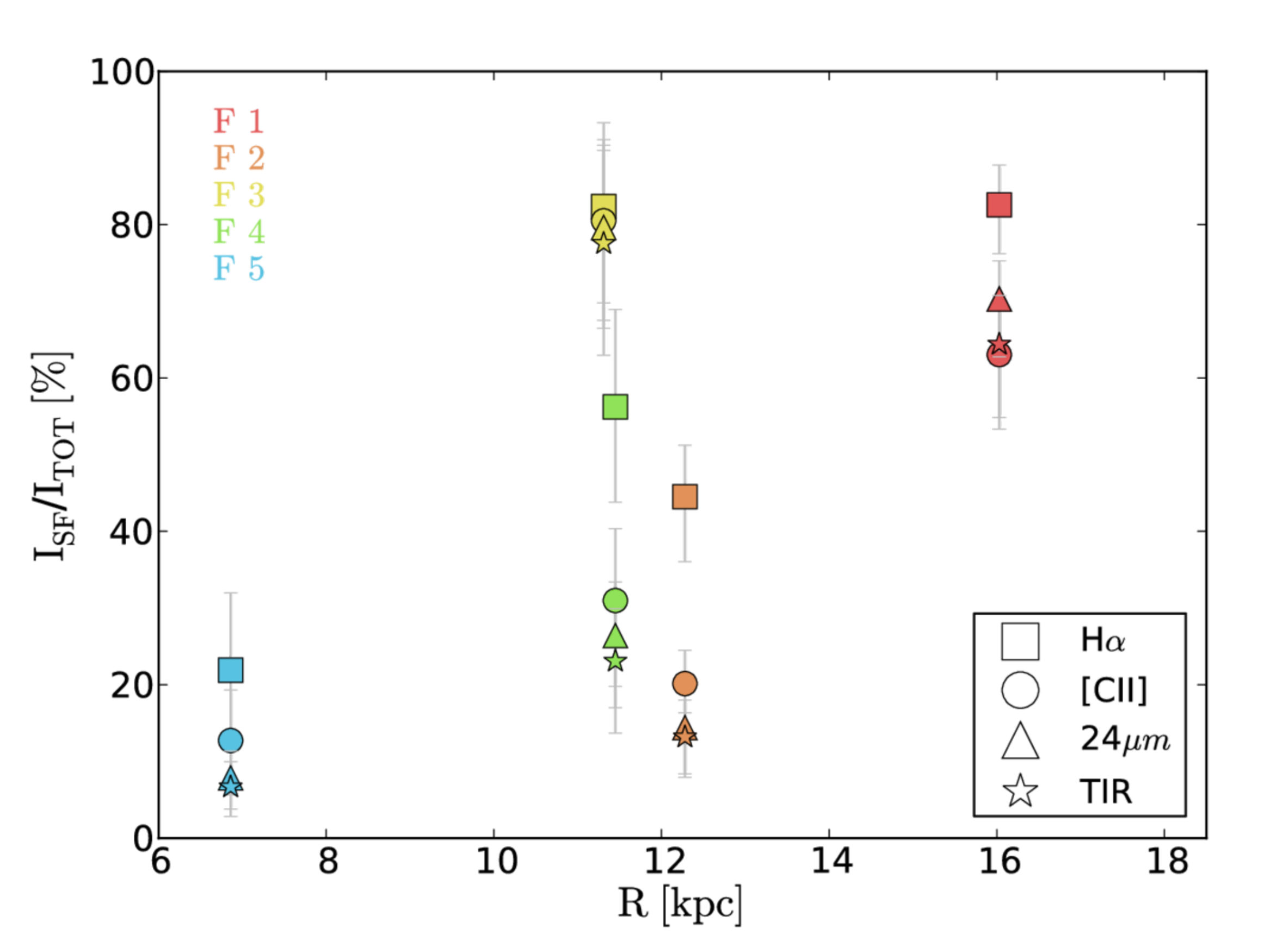}
\caption{Emission of H$\alpha$, [\cii], 24\mum\ and TIR from the SF region vs. the total emission in a given Field as a function of galactocentric radius. Error bars translate uncertainties in the H$\alpha$ contour level used to define ``SF regions'' (see \S~\ref{sec:result_reg}). Note, that if a 30\% higher H$\alpha$ level is picked, points of all tracers for a given Field will systematically shift down to the value indicated by a respective error bar.}
\label{fig:fracvsrad}
\end{center}
\end{figure}

To determine the emission fraction from ``SF regions'', ${\rm I_{SF}/I_{TOT}}$, we integrated the emission from each tracer within the contours and divided from the total emission in the map. The total emission only considered the area that had coverage in all relevant maps. In Figure~\ref{fig:fracvsrad}, we plot these estimates as a function of galactocentric radius.

\begin{figure}[!ht]
\begin{center}
\includegraphics[width=.5\textwidth]{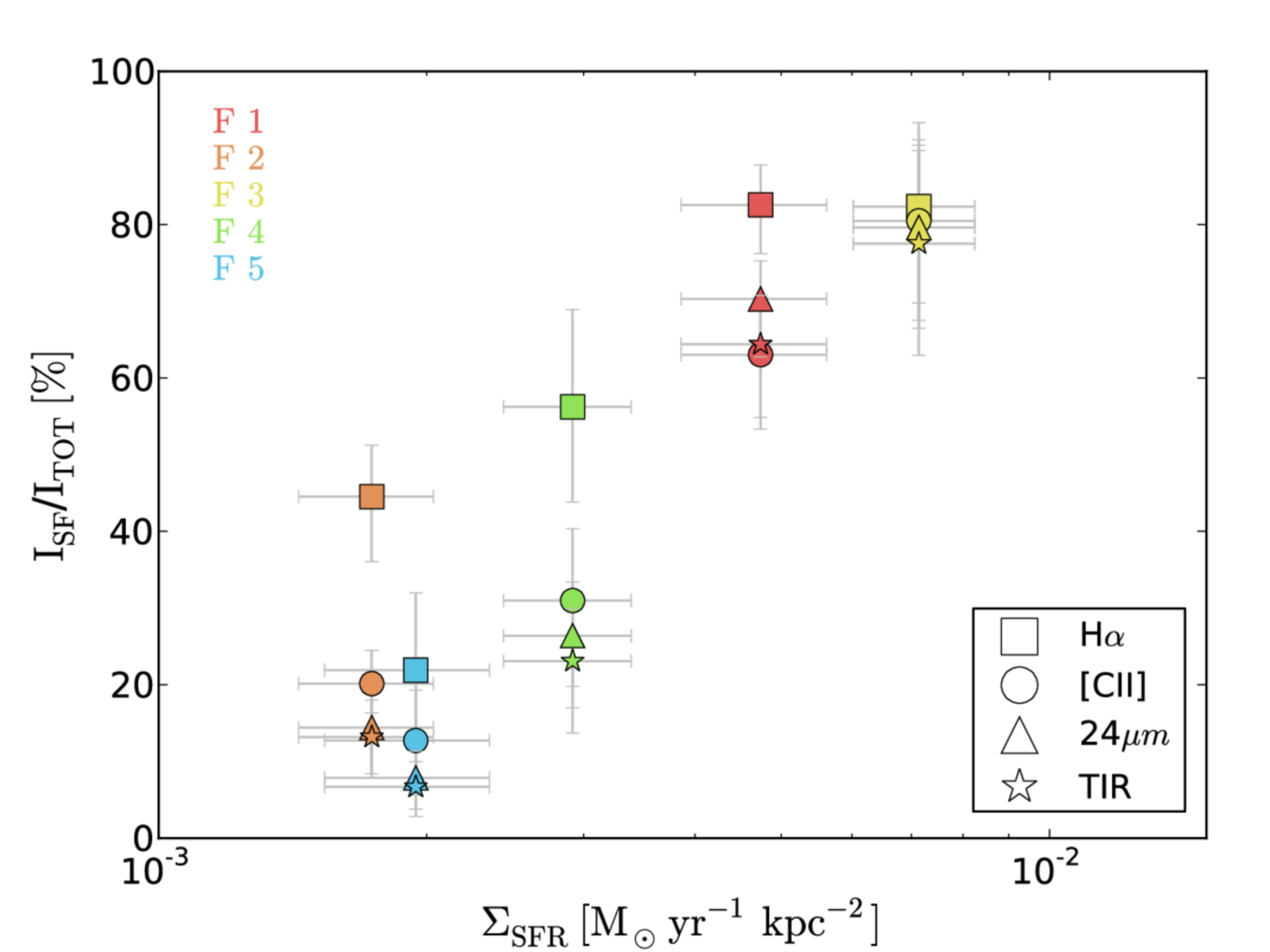}
\caption{Same emission fraction as in Figure \ref{fig:fracvsrad}, but instead of radial distance, fractions are plotted as a function of star formation rate surface density.}
\label{fig:frac}
\end{center}
\end{figure}

The fractions arising from SF regions varies systematically with the choice of tracer. H$\alpha$ always has the highest fraction of its emission arising from SF regions, as one might expect due to the regions' definition. The fractions of H$\alpha$ from SF regions range between $\sim$\,20--80\%.  
Previous studies have identified a large component of diffuse H$\alpha$ emission from M\,31 \citep[$\sim$\,63\%,][]{Azimlu2011} and we recover a similar value in Field 4.  On average our H$\alpha$ diffuse fraction is lower than 63\%, as our Fields are biased to high SFR surface densities by selection.

In general, TIR and 24\,\mum\ have similar fractions of their emission coming from SF regions, as shown in Figure~\ref{fig:fracvsrad}.  In most of the cases, they also have the lowest SF region fractions compared to the other tracers. The fraction of [\cii] from SF regions is intermediate between that of the IR tracers (TIR and 24\,\mum) and H$\alpha$, except in F1. 
In all of the cases [\cii] is closer to the IR than the H$\alpha$, aside from in F3 where all tracers give approximately the same fractions.
It is somewhat surprising that the 24\,\mum\ emission behaves similarly to TIR, since we expect it to trace warm dust dominantly heated by young stars whereas TIR traces all dust emission.  We discuss this finding further in Section~\ref{sec:disc}.

In Figure~\ref{fig:fracvsrad}, we explore the radial trends of the fractions of H$\alpha$, [\cii], TIR and 24\,\mum\ that arise from SF regions. Given that there are strong radial trends in metallicity, radiation field strength and stellar mass surface density in M\,31 we naively expect some trends, however no such radial trends are apparent. 
Fields 2, 3 and 4 lie at a similar galactocentric radius in the 10\,kpc ring, yet they span almost the whole range of SF fraction values. Given this, it is likely that local conditions are dominating the fractional contribution of SF in our Fields. The SFR and ISM
surface densities in M\,31 do not show radial trends, but rather are dominated by the spiral arm structure and the 10\,kpc ring \citep{Draine2014}. 

In Figure (\ref{fig:frac}), we find a much clearer trend of increasing fraction of H$\alpha$, [\cii], 24\mum\ and TIR from SF regions with increasing ${\rm \Sigma_{SFR}}$, with Fields 2, 3 and 4 following this trend. 
This demonstrates the dominance of the local conditions to the SF fractions, with the weak trend of SF fractions with radius in Figure~\ref{fig:fracvsrad} most likely driven by a combination of SFR, opacity and metallicity.

Even with these trends it is clear that there is always a substantial diffuse component to the [\cii] emission in M\,31, with even the Field with the highest ${\rm \Sigma_{SFR}}$ fraction, Field 3, showing a 20\% contribution to [\cii] from the diffuse phase.

Based on Figures~\ref{fig:fracvsrad} and \ref{fig:frac}, the main result is that [\cii] emission is more extended than H$\alpha$, but less extended than TIR and 24\,\mum.
We find that the large fractions of diffuse emission arising from outside the SF regions are anti-correlated with ${\rm \Sigma_{SFR}}$. We will discuss possible explanations in Section \ref{sec:disc}, including gas heating generated by a diffuse radiation Field and/or leaked photons from SF regions.

\subsection{The Correlation Between [\cii] and ${\rm \Sigma_{SFR}}$}
\label{sec:result_cii_sfr}

Due to its strength and accessibility at high redshifts with new sub-mm telescopes (e.g.~ALMA), the correlation between [\cii] and SFR has recently been explored in nearby galaxies \citep[within 200\,Mpc, e.g.][]{deLooze2014,Herrera2014}, to provide a calibration for this line. These studies have found a tight correlation of [\cii] surface brightness and $\Sigma_{\rm SFR}$ on kpc scales. However, from the previous section it is clear that even at these scales this will include contributions of diffuse emission to both the [\cii] line and the $\Sigma_{\rm SFR}$ measurement.

Using the high spatial resolution available in M\,31, we can investigate the correlation both on $\sim$\,50\,pc scales (our working resolution) and on $\sim$\,700\,pc scales \citep[i.e.~averaged over one full Field, matching the typical resolution of the KINGFISH galaxies in][]{Herrera2014}. 
On 50\,pc scales we can separate SF regions from diffuse emission as previously described in \S~\ref{sec:result_frac}.  By degrading our resolution to match nearby galaxy studies, we can investigate how the SF and diffuse components participate in creating the observed correlation between [\cii] and $\Sigma_{\rm SFR}$.

\subsubsection{The Correlation Between [\cii] and ${\rm \Sigma_{SFR}}$ on 50\,pc scales}
\label{sec:result_cii_sfr50}

\begin{figure*}[ht!]
\begin{center}
\includegraphics[width=1.\textwidth]{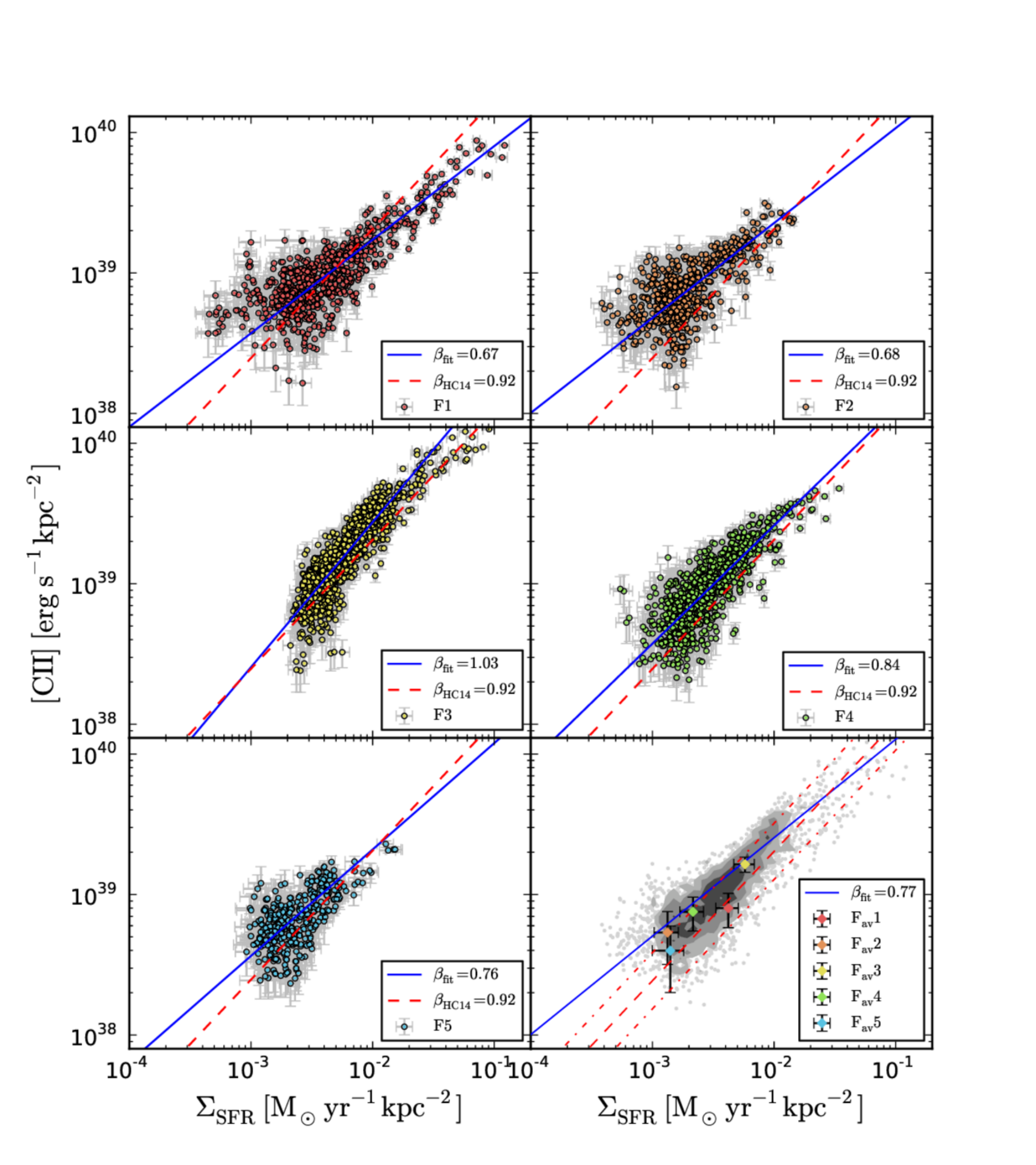}
\caption{[\cii] and SFR surface densities for individual 5.2\arcsec\ pixels ($\sim$\,20 pc physical scale) in all Fields. Only pixels with signal/noise per pixel greater than 3 in both measurements are included.  Note, that 3-$\sigma$ limit is approximate, because we use noise maps. The solid blue line shows the linear fit (in log space) to the data in each Field. The red dashed line is the relation determined by \citet{Herrera2014} for the full KINGFISH sample using data of $\sim$\,1\,kpc resolution (with  $\rm{\beta_{HC14}=0.94}$). Lower right panel: This figure reveals the [\cii] and SFR surface density relation from all Fields together (grey density plot and grey dots), as well as the relation averaged over whole Fields (diamonds, $\sim$\,700\,pc scales).  Note that number density plot, with grey levels 5-10, 10-20, 20-41 per bin size 0.05 dex, does not represent uncertainties. The blue line is the best fit to the individual pixels from all Fields together. The red dashed-dotted lines are 1$\sigma$ 
scatter around \citet{Herrera2014}'s fit. \\ \vspace{0.5cm}}
\label{fig:ciisfr50}
\end{center}
\end{figure*}

We first investigate the correlation between [\cii] and SFR for individual pixels in each of our Fields. The results are presented in Figure~\ref{fig:ciisfr50}. Each pixel has a physical scale of $\sim$\,20\,pc, meaning that we are slightly oversampling our physical resolution of $\sim$\,50\,pc. All pixels have $3\sigma$ significance in both the [\cii] and $\Sigma_{\rm SFR}$ measurements. Note that at these physical scales, $\Sigma_{\rm SFR}$ is not truly indicative of the underlying star formation rate. Rather it represents the average H${\rm \alpha+ mid}$-IR flux in each pixel, which could arise from both stellar populations intrinsic to the pixel and photons leaked from stellar populations in nearby regions \citep[][]{Calzetti2007}.

A clear correlation between [\cii] and $\Sigma_{\rm SFR}$ is already visible in each of our Fields, with this supported by the Pearson's correlation coefficients presented in Table~\ref{tab:coeff}. To quantify this relation, we use orthogonal distance regression, which allows us to fit a linear function to data points (in logarithmic space) simultaneously accounting for both the $x$ and $y$ errors. Our fits show sublinear to linear relations of [\cii] to ${\rm \Sigma_{SFR}}$, with the most linear slope found for F3 which has the highest ${\rm \left< \Sigma_{SFR} \right>}$. Details of the fits for all Fields are summarized in Table~\ref{tab:coeff}, and the best fit linear relation for each Field is overplotted (blue solid line) in Figure~\ref{fig:ciisfr50}. We have also included in each Field the fit determined by \citet{Herrera2014} for their integrated sample (dashed red line).

From Figure~\ref{fig:ciisfr50} we see that F1 and F2 have the flattest slopes, with both Fields showing an excess of [\cii] for the lowest ${\rm \Sigma_{SFR}}$. Note however that at [\cii] $\sim\,1.76 \times 10^{38}$ erg s$^{-1}$ kpc$^{-2}$ we hit the $1 \sigma$ sensitivity limit, and thus this excess may be due to a larger dispersion of [\cii] around ${\rm \Sigma_{SFR}}$, rather than a excess due to diffuse [\cii] emission.
Fields 3, 4 and 5 are consistent with the \citet{Herrera2014} trend, though F5 probes only a relatively small range in ${\rm \Sigma_{SFR}}$.

The errors of the \citet{Herrera2014} slope and intercept of the fit, where SFR is based on H$\alpha$ and 24\,\mum, are $\beta_{HC} = 0.8970\pm0.0078$, $\gamma_{HC} = 41.133\pm0.015$ (priv. com.). 
Our slopes are significantly flatter at a greater than 2-$\sigma$ level, except F3 which is significantly steeper. 

In general, the flatter slopes are consistent with the results in the previous section, as these slopes indicate a greater [\cii] fraction at low ${\rm \Sigma_{SFR}}$ (and hence low H$\alpha$ surface brightnesses and more ``diffuse'' regions). However, the flat slopes indicate that, while there is still a correlation between [\cii] and ${\rm \Sigma_{SFR}}$, the contribution of diffuse emission means that the calibration determined by \citet{deLooze2014} and \citet{Herrera2014} do not hold on these scales.

\begin{deluxetable}{ccccc}
\tablewidth{0pt}
\tablecolumns{5}
\tablecaption{The average [\cii] emission, $\Sigma_{\rm SFR}$ and metallicity of the Fields}
\tablehead{\multicolumn{5}{c}{$\log_{10}(\Sigma_{\rm [CII]}) = \beta \ \log_{10}(\Sigma_{\rm SFR}) + \gamma$ } \\
\hline
\multicolumn{1}{c}{} & 
\multicolumn{1}{c}{Pearson's} & 
\multicolumn{1}{c}{} & 
\multicolumn{1}{c}{} &
\multicolumn{1}{c}{} \\
\multicolumn{1}{c}{Field} & 
\multicolumn{1}{c}{correlation} & 
\multicolumn{1}{c}{$\beta$} & 
\multicolumn{1}{c}{$\gamma$} &
\multicolumn{1}{c}{$\sigma$\tablenotemark{b}} \\
\multicolumn{1}{c}{} & 
\multicolumn{1}{c}{coefficient $r$\tablenotemark{a}} & 
\multicolumn{1}{c}{} &
\multicolumn{1}{c}{} &
\multicolumn{1}{c}{[dex]} }
\startdata
1             & 0.80        & 0.67 $\pm$ 0.010 & 40.57 $\pm$ 0.019 &      0.17      \\
2             & 0.68        & 0.68 $\pm$ 0.015 & 40.71 $\pm$ 0.037 &      0.17      \\
3             & 0.89        & 1.03 $\pm$ 0.014 & 41.50 $\pm$ 0.031 &      0.13      \\
4             & 0.80        & 0.84 $\pm$ 0.016 & 41.09 $\pm$ 0.037 &      0.15      \\
5             & 0.64        & 0.76 $\pm$ 0.033 & 40.83 $\pm$ 0.084 &      0.16      \\
\hline
F$\rm{_{all}}$& 0.82        & 0.77 $\pm$ 0.009 & 40.84 $\pm$ 0.023 &      0.18    
\enddata
\label{tab:coeff}
\tablenotetext{a}{$r=1$ value of the Pearson's correlation coefficient indicates perfect correlation, on the contrary $r=0$ means no correlation. We found p-values extremely low, demonstrating that the correlation coefficients are significant. We found the highest p-value for F5 ($\sim10^{-41}$).}
\tablenotetext{b}{the dispersion of data points around the fit}
\end{deluxetable}

\subsubsection{[\cii] vs ${\rm \Sigma_{SFR}}$ on $\sim$\,700\,pc scales}
\label{sec:result_cii_sfr700}

To match the physical scales probed in the galaxy sample of \citet{Herrera2014}, we average our data in each Field to obtain the [\cii] surface brightness and SFR surface density at $\sim$\,700\,pc scales. 
We also attempt to match the removal of the cirrus contribution to SFR as done in \citet{Herrera2014}. The cirrus is the ``diffuse'' contribution to the total 24\,\mum\ emission caused by the heating of the dust by the interstellar radiation field  arising from older stellar populations not associated with the recent star formation. This cirrus leads to an overestimation of the $\Sigma_{\rm SFR}$. \citet{Leroy2012} determined the fraction of cirrus in nearby galaxies on kpc scales, using both physical modeling of the IR spectral energy distribution to determine the local radiation field and maps of the HI to determine the diffuse gas fraction. They found that the cirrus fraction decreases with increasing $\Sigma_{\rm SFR}$, and is small above ${\rm 10^{-2}\ M_{\odot} yr^{-1} kpc^{-2}}$. However, below this value, they found the contribution of the cirrus to 24\,\mum\ flux could be significant, albeit with a large scatter. As all of points are below this limit, we subtract a constant fraction of 24\,\mum\
 emission, $f_{cir}=0.4$, which is the average $f_{cir}$ in the range of $\Sigma_{\rm SFR}$ that we probe determined by \citet{Leroy2012}\footnote{ We do not to remove the cirrus on our smaller 50\,pc observations, as we are beginning to resolve out the ``diffuse'' emission of 24\,\mum\ on these scales. The cirrus correction of \citet{Leroy2012} was derived on scales of $\sim$\,1\,kpc, and cannot be simply applied to these smaller scales. In any case, due to our 3-$\sigma$ cut, H$\alpha$ is the dominant contributor to the SFR we measure for most of the pixels on these scales, such that correcting for the cirrus does not make a significant difference}. 
The value we use is higher than the median $f_{cir}=0.17$ used by \citet{Herrera2014}, however they probe higher $\Sigma_{\rm SFR}$ values on average, and we wish to assume a conservative correction here.
Our value of $f_{cir}$ is consistent with the high ``non-SF'' region contribution to the 24\,\mum\ we find in  Figure~\ref{fig:frac}. Removal of the cirrus makes the greatest difference to fields with low $\Sigma_{\rm SFR}$ (Fields 2, 4, and 5; $\sim 26\%$), and smaller differences to the higher $\Sigma_{\rm SFR}$ fields ($\sim 15\%$ for Fields 1 and 3). Nevertheless, we find that the details of the cirrus removal do not have a large impact on our results described below.

In the lower right panel of Figure~\ref{fig:ciisfr50}, we show the average value for each Field overlayed on the greyscale density plot for the individual pixels from all Fields combined. We repeat the orthogonal distance regression on the pixels from all Fields and have overplotted the fit (solid blue line), with the fit results also presented in Table~\ref{tab:coeff}. While the fit to all pixels shows a similar flat slope to the individual Fields ($\beta_{\rm all}=0.79$)\footnote{ Note, that the fit (blue line) does not go through the densest part of the number density plot, because the latter does not represent the uncertainties of the points, while the ODR fitting method takes them into account.}, we find that once averaged to $\sim$\,700\,pc scales, the data are consistent with the \citet{Herrera2014} relation within uncertainties. 
Therefore, while the [\cii] vs SFR relation is flatter at small physical scales ($\sim 50$\,pc) due to the separation of diffuse [\cii] emitting regions from the star forming regions, on kpc scales this diffuse emission is correlated sufficiently with star formation to provide the observed linear relation.

\subsection{Far-IR line deficit}
\label{sec:fir_def}

Another factor that affects the relation between [\cii] emission and SFR is the efficiency of converting absorbed photons from the star forming region to [\cii] emission.
Depending on whether the gas is ionized or neutral, this efficiency will differ as a result of different heating processes  (gas photoionization versus PE effect on dust grains).
In our regions overall, we estimate the fraction of [\cii] coming from ionized gas to be small. One region in Field 5 may have a larger than usual contribution from ionized gas, and is discussed in Section~\ref{sec:fir_def50}.

To get a rough estimate on the ionized [\cii] contribution we first make use of several locally significant [\nii]$\lambda 122\,\mu {\rm m}$ measurements in Fields 1 and 3. The line is detected only in the brightest regions, with local $\log \Sigma_{\rm SFR} > -1.5$. For these bright SF regions, we can estimate the ionized contribution to the total emission, [\cii]$_{\rm ion}$/[\cii], by following the prescription in \citet[][]{Croxall2012}, particularly their Figure 11. As the [\cii]/[\nii] ratio is density dependent and we have no constraint on the density, we assume a low density gas of $ n_{\rm e} \sim 2 \ {\rm cm^{-3}}$, resulting in a high value of [\cii]$_{\rm ion}$/[\nii]$ = 6$. Even with this assumption that provides us with an upper limit on the [\cii]$_{\rm ion}$/[\cii] ratio and probing the highest SF regions, we still find the ionized contribution to [\cii] to be less than 50\%, with [\cii]$_{\rm ion}$/[\cii]$ = 28\%$ and 40\% in the SF regions of Field 1 and Field 3, respectively. 

To further support the low [\cii]$_{\rm ion}$/[\cii] fraction, we estimate the contribution for the Fields as whole by using H$\alpha$ as a proxy. 
\citet{Groves2010} models predict [\cii]$_{\rm ion}$/H$\alpha \sim 0.05-0.1$ in \hii\ regions,  assuming relative efficiency of ionized gas cooling by [\cii] and H$\alpha$ lines. We measure everywhere the ratios of the surface brightness [\cii]/H$\alpha > 0.5$. 
Thus, we infer from the models [\cii]$_{\rm ion}$/[\cii]$ \sim 20\%$ (for [\cii]$_{\rm ion}$/H$\alpha=10\%$). However, we caution that our H$\alpha$ values were not corrected for extinction, and therefore we might slightly underestimate [\cii]$_{\rm ion}$ contribution. Nevertheless, we expect 10-50\% of the [\cii] could be from ionized gas, emphasizing that the majority of the [\cii] arises from neutral gas.

Therefore in M\,31 most of the neutral gas cooling occurs through the [\cii] emission line, due to the weak [\oi] detection and low [\cii]$_{\rm ion}$ fraction. Given this, and that the neutral gas is predominantly heated by the photoelectric effect on dust grains, the ratio of [\cii] to the total IR emission (TIR, tracing total dust absorption) should trace the photoelectric (PE) heating efficiency.

As described in the Introduction, [\cii]/TIR has been found to vary globally with galaxy properties such as the total IR luminosity. In particular, a decreasing trend of [\cii]/TIR versus $\nu I_{\nu} (70 \mu m)/\nu I_{\nu} (100 \mu m)$ (a proxy for the dust temperature), has been observed.  Based on the study of global measurements of 60 normal galaxies, [\cii]/TIR  was found to decrease at high dust color values \citep[i.e. warm dust temperatures;][]{Malhotra2001}, but it first was referred to as the FIR-line deficit for [\cii]/TIR falling below $10^{-3}$ at high dust color values by \citet{Helou2001}\footnote{As the FIR-line deficit definition is vague, hereafter, we adopt the one given by \citet{Helou2001}}.
Generally, [\cii]/TIR was found to be approximately constant (with some scatter) at low dust temperatures, but sharply decreasing in galaxies with warmer dust colors \citep[$\nu I_{\nu} (70 \mu m)/\nu I_{\nu} (100 \mu m)\gtrsim$ 0.95,][]{Croxall2012}. One of the most commonly given explanations for this deficit is grain charging, where warmer dust is more highly charged, increasing its PE threshold and thus decreasing the average energy returned to the gas per photon absorbed.
Most of the studies of the deficit have concentrated on the [\cii] line, because it is typically the brightest, but other lines have also been investigated, and show a similar deficit \citep[e.g. ${\rm [}$\oi${\rm], [O}$\,{\sc iii}${\rm], [}$\nii${\rm]}$;][]{GraciaCarpio2011}.

In our Fields, we already have hints that the  [\cii]/TIR ratio and thus PE efficiency is changing between our SF regions and more diffuse regions, with Figures \ref{fig:fracvsrad} and \ref{fig:frac} revealing that the TIR fraction from the diffuse gas is greater than that of [\cii] in all fields. Thus, [\cii]/TIR seems actually higher in SF regions, somewhat contrary to the results in previous works for galaxies as a whole.
However, we can also explore [\cii]/TIR versus dust colors using our higher resolution at $\sim$\,50\,pc scales.

\subsubsection{Far-IR line deficit on $\sim$\,50\,pc scales}
\label{sec:fir_def50}

\begin{figure*}[ht!]
\begin{center}
\includegraphics[width=1.\textwidth]{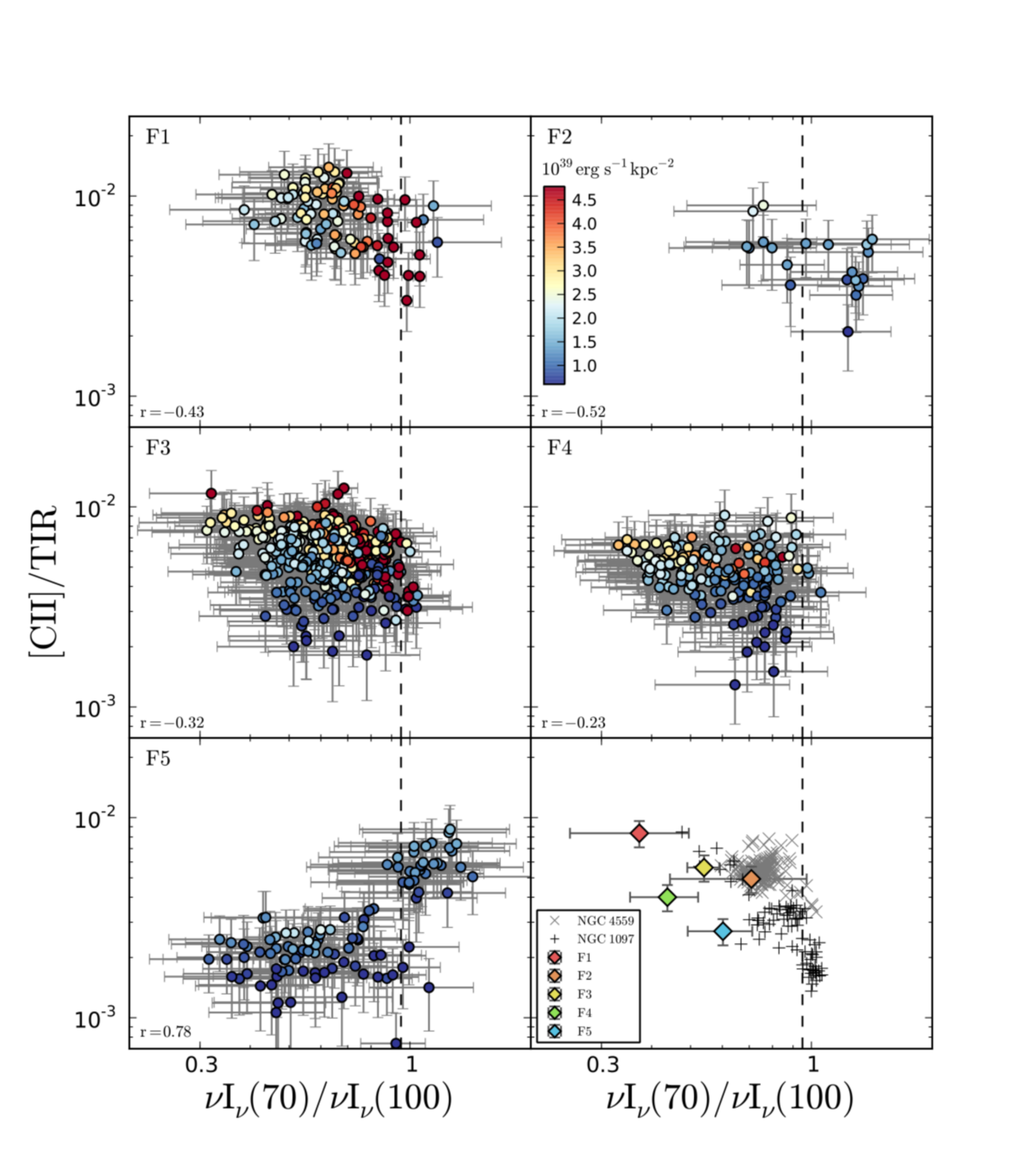}
\caption{[\cii]/TIR surface brightness ratio (proxy for gas heating efficiency) versus $\nu I_{\nu}(70 \mu m)$/ $\nu I_{\nu}(100 \mu m)$ (a proxy for dust temperature) for individual 5.2\arcsec pixels ($\sim$\,20\,pc physical scale) for each Field (as labelled in top left corners of each panel). Only pixels with S/N above 3-$\sigma$ in all quantities are shown, typically the 70\,\mum\ flux is the limiting quantity. 
The pixels are color coded by the [\cii] surface brightness (with respect to the scale given in the top right panel). In the lower left of each panel we indicate the Pearson correlation coefficient $r$ is given for each Field. $p$-values (roughly indicate the probability of an uncorrelated system) are high for F2 and F4, therefore correlation estimates for these Fields are not significant, while for the other Fields are sufficiently low to indicate a significant correlation.   
Lower right panel shows the same figure from \citet{Croxall2012} for two galaxies with $\sim$\,kpc pixels; NGC\,4559 (grey crosses) and NGC\,1097 (black '$+$' signs). The large diamonds show the integrated results for each of our Fields at the same physical scale ($\sim$\,700\,pc), color coded by the Field number. Note, that the averages on the last panel take all the measurements in each Field, not only the 3-$\sigma$ significant points.  \\ \vspace{0.5cm}}
\label{fig:firdef50}
\end{center}
\end{figure*}

In Figure~\ref{fig:firdef50}, we show the [\cii]/TIR surface brightness ratio versus the $\nu I_\nu(70 \mu m)$/ $\nu I_\nu(100 \mu m)$ ratio for individual pixels in Fields 1 to 5. Each pixel measures a region of $\sim$\,20\,pc, which slightly oversample our physical resolution of $\sim$\,50\,pc. Only pixels with a signal-to-noise greater than 3 in all quantities are plotted, with the 70\,$\mu$m flux being generally the most limiting quantity at these scales. In Figure~\ref{fig:firdef50}, all points are color coded by their [\cii] surface brightness as a proxy for diffuseness, where bluer pixels represent the more diffuse regions. Note that due to the limit of the 70\,$\mu$m flux, we do not probe the most diffuse regions explored in section \ref{sec:result_cii_sfreg}, as one can infer from the lower limit of the colorbar in the top right panel of Figure~\ref{fig:firdef50}.  

The [\cii]/TIR ratios in all our fields span approximately an order of magnitude, between $\sim 2\times 10^{-2}-10^{-3}$,  and span $\sim 0.3-2$ in dust color $\nu I_\nu(70 \mu m)$/ $\nu I_\nu(100 \mu m)$, a very similar phase space to where the bulk part of \citet{Malhotra2001} measurements for individual galaxies reside.  All of our observed [\cii]/TIR ratios lie above the $10^{-3}$ value which classically defines the `[\cii]-deficient' objects \citet{Helou2001}, thus we could state here that there is no [\cii] deficit regions across our fields in M\,31. The problem is that this classical deficit definition, as vague as it already is, was applied for global measurements, not for $\sim$\,50\,pc scales, at which scales it breaks down (manifested as a huge scatter in panels 1--5 in Figure~\ref{fig:firdef50}), due to different scale lengths (spatial distributions) of [\cii] and TIR (see section~\ref{sec:result_frac}). 

However, we do find moderate to weak negative correlations of the [\cii]/TIR surface brightness ratio with the $\nu I_\nu(70 \mu m)$/ $\nu I_\nu(100 \mu m)$ dust color in Fields 1 to 4 from the Pearson's coefficients' values ranging  from -0.23 to -0.52 (see Figure~\ref{fig:firdef50}). These weak correlations are consistent with the trends measured on global scales in previous studies, with weak negative correlations \citep{Malhotra2001} or constant [\cii]/TIR values found across 0.3--1 in dust color \citep{Helou2001,GraciaCarpio2011}. 

However, contrary to the other fields, we observe a very strong positive relation in Field 5, with Pearson's coefficient 0.78. This strong correlation is driven by the cluster of high [\cii]/TIR $\sim$\,0.006 points at the IR color of $\nu I_\nu(70 \mu m)$/ $\nu I_\nu(100 \mu m) \sim 1.2$, visible in the bottom left panel of Figure~\ref{fig:firdef50}. All these pixels are located in the star forming region in the south-east in Field 5. This bright region contributes on its own to $\sim$\,25-30\% of the total [\cii] emission in this $\sim$\,700\,pc Field. 
Despite a warm dust color, this region presents very weak IR emission, suggesting a low total dust mass. Given this low dust mass but relatively high [\cii] emission, it is likely that this region is dominated by \hii\ gas. Therefore, most of the [\cii] emission arises from photoioized gas, and not from neutral gas where the PE effect dominates the gas heating, which leads to the abnormally high [\cii]/TIR values for these dust colors.

At a given dust color, it is clear that we see a large spread of the [\cii]/TIR ratio, as great as any trends inferred over our observed IR color range. Interestingly, it is noticeable in all Fields that, at a given dust color, there is an increasing trend of [\cii]/TIR with [\cii] surface brightness. This clearly indicates that factors other than the dust temperature affect the PE efficiency. The details in each Field are sensitive to the geometry of the stars, dust and different phases of the gas.
One possibility is that at this resolution, we can see the effects of softer radiation fields in non-SF regions, such that there are sufficient photons to heat the dust to the observed colors, but fewer photons of sufficient energy to eject electrons from the dust grains, leads to a relatively cooler ISM in the diffuse phase.

\subsubsection{Far-IR line deficit on $\sim$\,700\,pc scales}
\label{sec:fir_def700}

In the lower right panel of Figure~\ref{fig:firdef50} we plot the results from \citet{Croxall2012} showing the [\cii]/TIR surface brightness ratio versus the $\nu I_\nu(70 \mu m)$/ $\nu I_\nu(100 \mu m)$ ratio on $\sim700$\,pc scales for two nearby galaxies: NGC\,4559 and NGC\,1097. 
We used the same instruments, data reduction and methodology (i.e. TIR prescription) as in \citet{Croxall2012} paper, which makes the comparison straight forward. Their resolution element scales are similar to our integrated Fields. The galaxies in their paper represent clear examples of a galaxy without a FIR-line deficit (NGC\,4559) and with a deficit (NGC\,1097).
These galaxies show the same range in both axes as we see with our higher physical resolution, but show a stronger trend of decreasing  [\cii]/TIR with IR color, with \citet{Croxall2012} clearly demonstrating a lower ratio for these objects with warmer colors.  On top of the results for these galaxies, we plot the integrated measurements for each of the Fields matching the  $\sim$\,700\,pc pixel sizes. 
This comparison shows that despite observing cooler dust colors in our Fields than those observed in \citet{Croxall2012}, we see a similar range in the [\cii]/TIR ratio as in both NGC\,4559 and NGC\,1097. However, it is clear that F1 has a higher than average ratio.

\begin{figure}[ht!]
\begin{center}
\includegraphics[width=.5\textwidth]{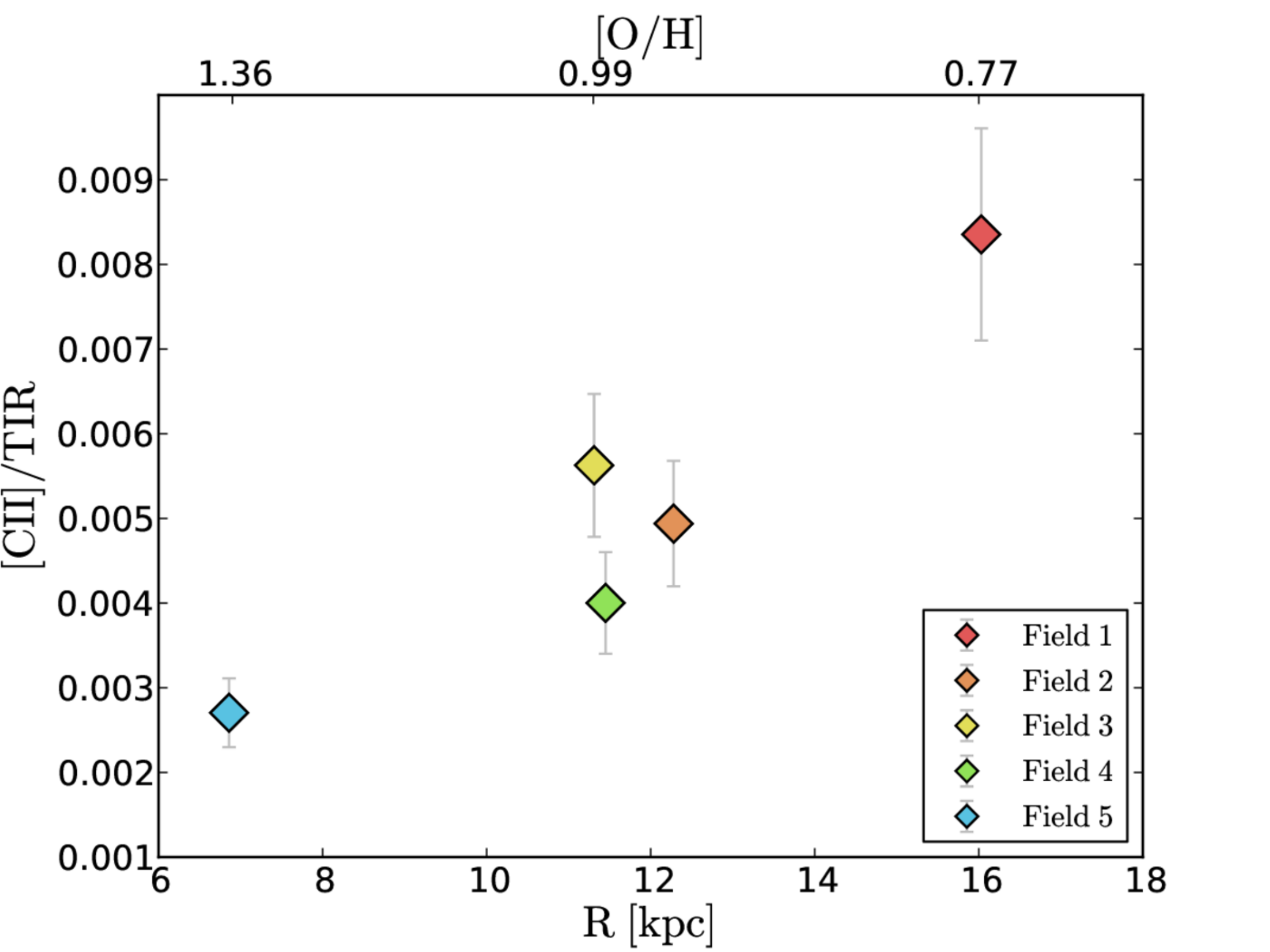}
\caption{The integrated [\cii]/TIR surface brightness ratio versus the galactic radius
(bottom axis) and average gas-phase metallicity (top axis; oxygen abundance as a proxy for metallicity in units relative to solar ${\rm (O/H)/(O/H)_{\odot}}$) for our 5 Fields. Fields 2 -- 4 lie in the same 10\,kpc gas rich ring, and have similar [\cii]/TIR values.}
\label{fig:cii_metal}
\end{center}
\end{figure}

With only 5 data points, we cannot conclusively identify a trend in the [\cii]/TIR ratio with IR color on these scales in M\,31. However, on closer examination, there is a trend with radius, shown in Figure~\ref{fig:cii_metal}.
We observe in Figure~\ref{fig:cii_metal} a strong radial relation with the [\cii]/TIR ratio, with a lower value presented in the inner region, similar ratios seen for Fields 2 to 4 which all lie in the same 10\,kpc gas rich ring that dominates IR and ISM images of M\,31. 
A similar radial trend of the [\cii]/TIR ratio is seen in M\,33 \citep{Kramer2013}. Smith et al. (in prep.) find a similar decreasing relation as a function of the gas-phase metallicity on $\sim$\,1\,kpc scales using the much larger sample of KINGFISH galaxies.

However, not only the metallicity varies with radius in M\,31, but many properties vary across M\,31 and could possibly affect the [\cii]/TIR ratio, such as the stellar density and radiation field strength, as already seen at the higher resolution. We discuss these, and the association with the diffuse and SF regions in the following section.

\section{Discussion}
\label{sec:disc}

As the nearest massive spiral galaxy, M\,31 is an ideal laboratory to study the ISM on small scales while still being comparable to studies of similar galaxies at larger distances. Making use of this high physical resolution, we have explored the origins of [\cii] and the dominant ISM heating processes by comparing the line emission strength and distribution against other tracers of ISM heating, the H$\alpha$ line, and the 24\mum\ and TIR continuum.

\subsection{[\cii] on $\sim$\,50\,pc scales}
\label{sec:disc50}

At our limiting resolution of $11\arcsec$ ($\sim$\,50\,pc), we clearly see a strong correlation of the [\cii] emission with H$\alpha$ through the spatial coincidence of the surface brightness peaks in the maps in Figures~\ref{fig:ha_cii_maps} and~\ref{fig:ha_cii_maps2}. 
However, the [\cii] emission in the maps appear to be more spatially extended than the H$\alpha$. We quantified this spatially extended [\cii] emission by separating each Field into ``SF regions'' and ``diffuse'', based on a H$\alpha$ surface brightness cut that encompassed the majority of \hii\ regions. We find that [\cii] is always more spatially extended than H$\alpha$, which arises predominantly from ``SF regions'', but less so than the dust emission (traced via the TIR), which had the highest diffuse fraction in all Fields. As our ``SF regions'' are 100's of pc in physical size they encompass several \hii\ regions (as seen in Figures ~\ref{fig:ha_cii_maps} and~\ref{fig:ha_cii_maps2}), thus the larger diffuse fractions cannot simply be the result of us resolving out the \hii\ regions and associated photodissociation regions, but indicative of a real diffuse emission. 

The association of [\cii] with recent star formation is further supported by the significant correlation between the [\cii] surface brightness and SFR surface density on $\sim$\,50\,pc scales (Figure \ref{fig:ciisfr50}). The presence of a diffuse contribution to the [\cii] emission is indicated by the sub-linear slopes found for all Fields in the logarithmic relationship.

Thus the resolved [\cii] emission in Andromeda reveals that some fraction of the [\cii] line originates from gas directly heated by UV photons from star-forming regions, with this fraction dependent upon the local SFR surface density. While this SF region [\cii] includes both contribution from the ionized gas in \hii\ regions and the neutral gas in the directly associated photodissociation regions which are unresolved in our observations, we find that the \hii\ region contribution must be less than 40\% across our map based on our few [\nii] 122\mum\ detections and model-based estimates from the H$\alpha$ maps. The remaining [\cii] flux arises from a diffuse phase, not spatially coincident with star-forming regions, that is also associated with a large fraction of the TIR surface brightness.

The heating source for this diffuse phase is not directly obvious. We consider two mechanisms for the heating of this diffuse phase: (1) photon leakage from SF regions, and (2) a distinct diffuse UV radiation field. 

In the first mechanism, the diffuse gas and dust are heated primarily by photons that have escaped the immediate vicinity of the young, massive stars that emit them. Such a mechanism has been put forward to explain the diffuse H$\alpha$ emission. 
As the FUV photons that heat the dust and gas, are less energetic than ionizing photons and have a longer mean free path, will cause the spatial extent of the diffuse [\cii] emission to be greater than that of H$\alpha$. In this scenario, even though we have seen a large diffuse [\cii] component, it is still associated with the stars that heat SF regions. This means the use of [\cii] as a SFR indicator should still be valid on scales large enough ($\sim$kpc) to average over these leakage effects.
Possible evidence for this mechanism are the overdensity of points below the best-fit line at the bright end of the [\cii]-SFR relation in F3, F4 and F5 (Figure~\ref{fig:ciisfr50}) that could indicate regions where the ionizing photons have been absorbed, but the FUV photons have leaked to adjacent regions.

The second mechanism is where the diffuse UV radiation field arises from sources different to the massive stars powering the SF regions. The most likely source of this diffuse UV field would be B-stars, as in the solar neighborhood \citep{Mathis1983}. B-stars generate sufficient far-UV photons to heat the neutral gas, but negligible amounts of ionizing photons.  In addition, they would have a more uniform spatial distribution than O-stars due to their longer lifetimes. If the [\cii] is predominantly heated by these stars, it will still measure the SFR, but over longer timescales than H$\alpha$. 

Both of the suggested mechanisms will result in a softer radiation field outside of the SF regions. If the heating photons come purely from recent star-formation, we expect a softening of the radiation field with distance from the SF regions as the harder photons are preferentially absorbed. If the diffuse UV radiation field arises from a different stellar population, a softening of the radiation with distance from the SF regions requires that the mean age of the stellar population increases with distance from the youngest stars. 
Further evidence for the diffuse heating mechanisms can be seen in Figure~\ref{fig:firdef50}, where, in every Field, for a given dust color (i.e.~dust temperature) there is an increasing trend of [\cii]/TIR with increasing [\cii] surface brightness. This suggests that the brightest [\cii] emitting regions (which we have shown to be associated with SF regions) have a higher heating efficiency. One way to produce such a trend is to have softening of the radiation field from SF regions to the more diffuse ISM, reducing the relative energy input to the gas (requiring $>6$\,eV photons) as compared to the dust heating (which absorbs all photons).

Realistically, it is likely that both mechanisms play a role, with leaked FUV photons from SF regions gradually merging with the radiation field from an underlying diffuse stellar population. Only through the knowledge of the relative distribution of the stellar population in comparison with the observed dust emission could the relative contribution of each mechanism to the [\cii] emission across our Fields be disentangled.

\subsection{[\cii] on $\sim$\,700\,pc}
\label{sec:disc700}

It is clear from the previous section (\S~\ref{sec:disc50}) that at the larger scales of our integrated Fields ($\sim$\,700\,pc), there will be contributions from diffuse emission to both the [\cii] surface brightness and $\Sigma_{\rm SFR}$ (based on H$\alpha$ and 24\,\mum). Based on Figure~\ref{fig:frac}, this diffuse fraction contribution will be greater than 50\% in some regions, depending upon the local SFR surface density.
Nevertheless, when we look at these $\sim$\,kpc scales, we find that the integrated Fields are consistent with the [\cii]--$\Sigma_{\rm SFR}$ relation found on similar scales by \citet{Herrera2014}.  From this we can infer that this diffuse emission is sufficiently correlated with SFR on these scales that the relation holds, meaning that, for our two suggested mechanisms for the heating of the diffuse gas, on these scales we must capture all leaked photons or that the source of the diffuse radiation field, i.e.~B-stars, are co-located with star-formation on $\sim$kpc scales.

On the other hand, we do find large scale variations of [\cii]/TIR with galactocentric radius in M\,31 (Figure~\ref{fig:cii_metal}). If the [\cii]/TIR ratio is tracing PE heating efficiency, then it is puzzling how the relationship with SF can stay the same as in \citet{Herrera2014}, when this efficiency appears to vary so dramatically, by a factor of $\sim 3$, between Fields 1 and 5. Taken together what this suggests is that there is a constant FUV energy fraction emitted by stars  transferred to gas that is then cooled by [\cii], but that the TIR must decrease relative to both [\cii] and SFR with increasing galactocentric radius.

While a changing ionized gas contribution to the [\cii] emission could possibly explain the [\cii]/TIR ratio, we have demonstrated that this contribution is less than 50\% in even the highest $\Sigma_{\rm SFR}$ pixels at 50\,pc scales, and thus will be much less on the scales of our integrated Fields. Thus this cannot explain the factor of $\sim 3$ variation.

It is also possible that the fraction of the FUV energy absorbed by dust that goes into the gas can change (an increase in the physical photoelectric heating efficiency) and thus alter the [\cii]/TIR ratio. However this also requires some form of ``conspiracy'' to maintain the [\cii]--$\Sigma_{\rm SFR}$ relation, such as decreasing the FUV absorbed by dust relative to the SFR when the PE efficiency increases. Thus, while we cannot discount this possibility, we believe it to be unlikely. This is further supported by the non-monotonic radial trend in $q_{\rm PAH}$, the mass fraction of dust in PAHs, as determined by \citet{Draine2014}, in particular their Figure 11. A larger mass fraction of PAHs should lead to more efficient heating \citep{Bakes1998}, yet  the determined trend in $q_{\rm PAH}$ peaks at $\sim$11.2\,kpc, unlike the monotonic [\cii]/TIR ratio.

A more likely physical explanation for the [\cii]/TIR decrease towards galaxy center might lie in the fact that dust can be heated by both UV and optical photons, while both H$\alpha$ and [\cii] require harder photons ($>13.6$ and $\gtrsim$6\,eV, respectively). Thus to reduce the TIR relative to [\cii] as a function of radius, the FUV absorbed by dust must increase relative to the lower energy photons ($<6$\,eV) absorbed by dust. This is possible through changing either the intrinsic heating spectrum or by changing the dust properties.

The hardness of the intrinsic stellar spectrum heating the dust will change with both stellar metallicity and local star formation history (parameterized through the mean stellar age). The gas-phase metallicity, and presumably the stellar metallicity, decreases monotonically with radius, with a factor of 2 change between Field 1 and Field 5 \citep[based on the dust-to-gas ratios work in][]{Draine2014}. To assess the impact of metallicity on spectral hardness, we compare two almost identical $10^6\,M_{\odot}$ clusters at the age of 10\,Myr, modelled with {\tt Starburst99} \citep{Leitherer1999}, with an instantaneous star-fromation burst and Kroupa initial mass function, differing only by solar ($Z=0.02$) and subsolar (LMC, $Z=0.004$) metallicities. The relative hardness increases by $\sim15$\% from the solar to subsolar metallicity cluster, as measured by comparing the change in the FUV ($\nu > 6\,eV$) to NUV ratio.  While this will contribute to the observed trend in [\cii]/TIR, it is not sufficient to 
explain the factor of 3 change.

The local star formation history can strongly affect the heating spectrum seen by dust. A clear example of this is the work of \citet{Groves2012}, who showed that in the bulge of M\,31 an old stellar population dominates the heating radiation field, with its extremely soft radiation field meaning that optical light actually dominated the heating of dust. Thus we do expect a hardening of the radiation field with increasing radius in M\,31 as we move from older bulge dominated regions to younger disk dominated regions. Evidence for this can be seen in the observed radial FUV-NUV color in M\,31 shown in \citet{Thilker2005}, where the bluer color with radius may indicate a younger mean stellar age. The SFR, however, does not vary monotonically with radius, peaking in the 10\,kpc ring, and this will also affect the hardness of the local radiation field.

Changing the dust properties can also change the [\cii]/TIR ratio by changing the relative amount of FUV photons absorbed. One clear way is to change the dust absorption curve, which steepens with decreasing metallicity as seen through comparison of the extinction curves of the Milky Way versus the LMC \citep{Gordon2003}. The steepening from a Milky Way opacity to a LMC opacity, as plausibly expected from a decrease in the metallicity by a factor of 2, will mean relatively more FUV photons will be absorbed, however this increase is relatively small \citep[see Figure 10 in][]{Gordon2003} and will not contribute a significant amount to the [\cii]/TIR increase.

In addition to steepening the opacity curve, decreasing the metallicity will also decrease the dust-to-gas ratio (DGR), which has been found to decrease monotonically with radius in M\,31 \citep{Draine2014}. Decreasing the DGR will decrease the overall dust opacity for the same column of gas, increasing the mean free path of optical and UV photons, and decreasing the total amount of photons absorbed \citep[a similar argument was put forward by][for the high ratios observed in the LMC]{Israel1996}. Lower dust opacity will increase the average energy absorbed by dust, as preferentially the FUV photons are absorbed and the NUV-optical photons escape (due to the steep power-law nature of the dust opacity), and this decreasing opacity will lead to some part of the radial FUV-NUV color gradient observed in \citet{Thilker2005}. Supporting this idea, we find that the 24\mum/H$\alpha$ ratio (a measure of the SF region extinction) increases with decreasing [\cii]/TIR ratio, albeit in a non-linear manner.  However, the 
overall dust opacity depends upon the total dust column, and this has been found to peak around 
the 10\,kpc ring in M\,31 \citep{Draine2014}, and is the region with the highest expected extinction \citep{Tempel2010}. This non-monotonic trend in total extinction goes against the simple trend seen in Figure~\ref{fig:cii_metal}.

A similar explanation was suggested by \citet{Israel1996} for an increase of the [\cii] emission relative to IR that they found with decreasing metallicity in an exploration of the LMC. In the low metallicity environment of the LMC ($Z_{\rm LMC}=0.004$), [\cii]/FIR was found to be $\sim$10 times higher than found in Milky Way \citep[$Z_{\rm \odot}=0.02$; see ][and references therein]{Israel1996}. Their explanation for this was that at the low metallicities, the clumpy nature of the ISM caused deeper penetration of FUV photons into the molecular clouds, for the same A$_{\rm V}$, increasing the [\cii] flux for the same absorbed radiation.

Therefore, it is likely that a combination of a radial variation in both the dust opacity and star formation history causes the observed radial trend in the [\cii]/TIR ratio in M\,31.
While we favor the change of the radiation field due to mean stellar age as being primarily responsible for the observed radial trend in [\cii]/TIR due to the observed FUV-NUV and extinction gradients, to correctly disentangle which of the mechanisms dominates in M\,31 will require a determination of the spatially resolved star formation history in M\,31. This is currently being undertaken by the Pan-chromatic Hubble Andromeda Treasury Survey \citep[PHAT][]{Dalcanton2012,Lewis2014}, covering all of our Fields. With the intrinsic heating stellar spectrum being determined from this analysis, we will be able to correctly account for this effect on the [\cii]/TIR ratio and demonstrate whether this is indeed the dominant mechanism.  

For galaxies in the Local Group, we should be able to demonstrate the same effects, by comparing resolved star formation histories 
and extinction maps against the observed [\cii]/TIR ratio. This should answer to what extent the observed [\cii]/TIR trends in galaxies are due to heating effects or true changes in the photoelectric heating efficiency.

\section{Conclusions}
\label{sec:concl}

In this paper we present an analysis of [\cii] 158\,\mum\ emission in five Fields in M\,31. Combined with ancillary  H$\alpha$ and IR emission data, we studied the origins of [\cii], its relation with the SFR and the ISM properties. In particular, we have found:

\begin{itemize}

\item Significant amounts of [\cii] line emission are coming from outside the SF regions.

\item Even though we measure a large diffuse [\cii] fraction, integrated over $\sim$\,kpc scales, [\cii] still traces the SFR very similarly to what we see in larger samples of more distant galaxies. We explore different mechanisms that could be responsible for this diffuse phase including leakage of photons from \hii\ regions or diffuse UV radiation field generated by B stars. More diffuse [\cii] than H$\alpha$ emission is consistent with flatter slopes.

\item ${\rm [}$\cii${\rm ]}$ and SFR are correlated, but with a shallower slope than seen on $\sim$\,kpc scales.  This may be a result of the same diffuse [\cii] emission.

\item All of our observed [\cii]/TIR ratios lie above the $10^{-3}$ value, which classically defines the `[\cii]-deficient' objects. 
Yet this is not surprising as we explore much smaller scales than the global measurements that defined the deficit, and much more quiescent conditions than the centers of ULIRGs in which this deficit is clearly seen.
On 700\,pc our Fields do show a tentative decreasing trend of [\cii]/TIR with 70\,\mum/100\,\mum, however, with only 5 points, considerable scatter and large dust color uncertainties, it is not significant.
On the smaller 50\,pc scales we do generally see a weak correlation of decreasing [\cii]/TIR with warmer dust colors. However, this trend is inverted in F5 and in all fields we see a significant scatter ($\sim$\,order of magnitude) at a given dust color, that may be related to the [\cii] surface brightness.

\item We observe a large scale gradient of [\cii]/TIR across the disk of M\,31.  We explore potential causes for this trend and argue that a combination of effects due to changes in the dust-to-gas ratio, dust extinction curve, star formation history and radiation field are likely responsible.

\end{itemize}

Using [\cii] to trace the massive SFR, one must consider possible contributions to ISM gas heating by older stellar populations that can lead to tracing longer timescales, and/or leaked photons from \hii\ regions. The issue caused by the latter should go away when averaged over larger scales $\sim$\,few hundred pc. We will be able to shed some light in a following paper resolving stellar populations and their energy input in M\,31 using the PHAT survey (Kapala et al. in prep.).

\section*{Acknowledgements}

M. J. K. acknowledges funding support from the DLR through Grant 50 OR 1115.  K. S. acknowledges funding from a Marie Curie International Incoming Fellowship. The authors thank A.~Bolatto, R.~Herrara-Camus, J.~D.~Smith, H.-W.~Rix, S.~Glover, S.~Meidt, and A.~Hughes for helpful conversations in the course of this project.  We thank R.~Herrara-Camus for providing an early version of his paper for comparison. The authors would also like to thank the anonymous referee for providing us with very constructive comments.
This research made use of (1) Montage, funded by the National Aeronautics and Space Administration's Earth Science Technology Office, Computation Technologies Project, under Cooperative Agreement Number NCC5-626 between NASA and the California Institute of Technology. Montage is maintained by the NASA/IPAC Infrared Science Archive.
(2) the VizieR catalogue access tool, CDS, Strasbourg, France. The original description of the VizieR service was published in \citet{Ochsenbein2000}.
(3) the NASA/IPAC Extragalactic Database (NED) which is operated by the Jet Propulsion Laboratory, California Institute of Technology, under contract with the National Aeronautics and Space Administration. This research has made use of NASA’s Astrophysics Data System Bibliographic Services.
PACS has been developed by a consortium of institutes led by MPE (Germany) and including UVIE (Austria); KU Leuven, CSL, IMEC (Belgium); CEA, LAM (France); MPIA (Germany); INAF-IFSI/OAA/OAP/OAT, LENS, SISSA (Italy); IAC (Spain). This development has been supported by the funding agencies BMVIT (Austria), ESA-PRODEX (Belgium), CEA/CNES (France), DLR (Germany), ASI/INAF (Italy), and CICYT/MCYT (Spain).

\bibliographystyle{apj}
\bibliography{biblio}

\end{document}